# Progress, challenges and perspectives of computational studies on glassy superionic conductors for solid-state batteries

Zhenming Xu [a, *], Yongyao Xia [a, b]

a. *College of Materials Science and Technology, Nanjing University of Aeronautics and Astronautics, Nanjing 210016, China*

b. *Department of Chemistry and Shanghai Key Laboratory of Molecular Catalysis and Innovative Materials, Institute of New Energy, iChEm (Collaborative Innovation Center of Chemistry for Energy Materials), Fudan University, Shanghai 200433, China*

**Abstract:** Sulfide-based glasses and glass-ceramics showing high ionic conductivities and excellent mechanical properties are considered as promising solid-state electrolytes. Nowadays, the computational material techniques with the advantage of low research cost are being widely utilized for understanding, effectively screening and discovering of battery materials. In consideration of the rising importance and contributions of computational studying on the glassy SSE materials, here, this work summarizes the common computational methods utilized for studying the amorphous inorganic materials, reviews the recent progress in computational investigations of the lithium and sodium sulfide-type glasses for solid-state batteries, and outlines our understandings of the challenges and future perspective on them. This review would facilitate and accelerate the future computational screening and discovering more glassy-state SSE materials for the solid-state batteries.

[*]Corresponding authors:
E-mail: xuzhenming@nuaa.edu.cn

## 1. Introduction

The increasing demands for the large-scale energy-storage in many scenarios claim higher energy density and more safety of the rechargeable batteries beyond the current commercial lithium-ion batteries. The non-flammable solid-state electrolyte (SSE) materials to replace the liquid organic solvents and can effectively combine with lithium metal anode, simultaneously increasing the safety and energy density of all-solid-state batteries[1-5]. Thus, developing suitable solid-state electrolyte (SSE) for the alkali-metal ion batteries is of great significance, especially for lithium and sodium ion battery systems. So far, an increasing number of SSE materials have been developed to satisfy the wide applications of solid-state lithium and sodium ion batteries at room temperature, mainly including polymers[6-9], oxides[10-12], sulfides[13-18] and halides[19-22]. Among them, sulfide-based SSE materials have much higher ionic conductivities than those of organic polymers, oxides and halides[23-25]. Usually, sulfide-based SSEs can be divided into three categories according to their structural features, including crystalline phase, amorphous (glassy) phase, and glass-ceramic (partially crystalline) phase, most of which are well-known superionic conductors[14, 26]. Among the crystalline sulfide-based SSEs, $Li_{10}GeP_2S_{12}$(LGPS)[24] and $Li_7P_3S_{11}$[18], have the much higher room temperature ionic conductivities of 12 and 17 mS cm$^{-1}$ at room temperature (RT), respectively. In addition, another LGPS family material, the Cl-doped silicon-based $Li_{9.54}Si_{1.74}P_{1.44}S_{11.7}Cl_{0.3}$ shows the state-of-the-art ionic conductivity of 25 mS/cm at RT[13].

Interestingly, the amorphous sulfide-based SSEs, such as $Li_2S$-$P_2S_5$ and $Na_2S$-$P_2S_5$ glasses, show a better utilization in solid-state lithium (sodium) ion batteries compared to their corresponding glass-ceramics[27-28]. Glasses are superior to superionic ceramics in terms of their inherent isotropic ionic conduction, zero grain-boundary resistance, the good mechanical plasticity for compensating the unavoidable local mechanical pressures during cycling, and the chemical flexibility in terms of composition[29-30]. Glassy SSE materials are inherently metastable, in which the combined disordered short-range-order and long-range-order structures result in larger and more available free volumes and defects for ion conduction compared to the most stable crystalline phase with the same chemical composition[31-32]. In addition, lithium (sodium)-anion polyhedra in the glass phase are distorted to a certain extent compared to the crystalline structure, providing the activated lithium (sodium) ion occupations and frustrated energy landscapes for ion migration[33]. Additionally, $Li_3PS_4$ glass-ceramic solid electrolytes show better electrochemical stabilities in the all-solid-state $TiS_2$/SSE/$Li_{0.5}$In cells than LGPS solid electrolytes in spite of its superior ionic conductivity[34]. For the $Li_2S$-based glass, such as

$Li_2S–P_2S_5$[35-36], $Li_2S$-$P_2S_5$-$Li_2O$[37], $Li_2S$-$P_2S_5$-$SiS_2$[38], $Li_2S$-$P_2S_5$-$GeS_2$[39], $Li_2S$-$P_2S_5$-$P_2O_5$[40], $Li_2S$-$P_2S_5$-$LiI$[41] and $Li_2S$-$P_2S_5$-MS (M = Ca, Sr and Ba)[42], show relatively higher lithium ionic conductivities of ~ 0.1 mS/cm at RT. Among them, the lithium ionic conductivity at RT of the 75mol%$Li_2S$–25mol%$P_2S_5$ glass has been reported to be as high as 0.3-1 mS/cm [35, 43]. It's worth noting that the RT-phase $\gamma$-$Li_3PS_4$ crystallizing from the 75mol%$Li_2S$–25mol%$P_2S_5$ glass would significantly lower ionic conductivity of glass[44], while a certain amount of $\beta$-$Li_3PS_4$ crystallization (high-temperature phase) from the 75mol%$Li_2S$–25mol%$P_2S_5$ glass would still maintain the relatively higher ionic conductivity of glassy-ceramic sulfide solid electrolytes[45-47]. In addition, $Na_2S$–$P_2S_5$[48-50], $Na_2S$-$P_2S_5$-$SiS_2$[51] and $Na_2S$–$P_2S_5$–$B_2S_3$[52] glasses exhibit good sodium ionic conductivities, more than 0.1 mS/cm at RT.

With the rapid development of computing power, the computational material techniques with the advantage of low research cost are being widely utilized for understanding, effectively screening and discovering of metal-ion battery materials at the atomic scale[53-55], including cathodes[56-57], anodes[58-59], electrolytes[60-61], coating materials[62-63] and electrode/electrolyte interface[64-65]. The computational techniques greatly complement the experimental characterizations and significantly shorten the development cycle of a target material[66-67]. For computationally studying the important properties of an electrolyte material, such as the atomic structures, thermodynamics and ion diffusion kinetics, the density functional theory (DFT) calculations, kinetic Monte Carlo simulations, classical molecular dynamics (MD) simulations, and *ab*-initio molecular dynamics (AIMD) simulations are being widely applied in the battery communities[60-61, 68-69].

Considering the rising importance and contribution of computational studying on the research of glassy SSE materials, this review collects the common computational methods of studying the amorphous inorganic materials, summarizes the recent progress in the computational studying of the lithium and sodium sulfide-type glasses for solid-state batteries, provides our understandings of the challenges and potential future on them. Although the reported experimental and computational work of the glassy-state SSE materials are relatively fewer than crystalline, we believe that there are some chemical spaces for exploring the new glasses with the elementary composition of A-B-C-D (A = Li, Na and K, B = O, S, Se and Te, C is P, and D = Cl, Br and I). Therefore, the material computational researchers are amenable to screen and discover new glasses in the whole chemical space, but also should fully figure out the relationships among the structure, composition and ion diffusion property of glasses. We hope this review could facilitate and accelerate the future computational screening and discovering more glassy-state SSE materials for the solid-state batteries. This review is organized in the following pattern:

■ We first provided a comprehensive summary of the common computational methods that are utilized for studying the amorphous inorganic materials, including amorphous structure construction, structural properties and dynamical properties;

- Next, we made detailed reviews on the previous computational studies of the lithium and sodium sulfide-type glasses for solid-state batteries;

- Finally, we outlined our understandings of the challenges and potential future developments in the computational studying on the new glassy-state SSE materials for solid-state batteries.

## 2. Computational techniques

In this section, we summarize some technical details of the common computational methods that are used for studying the amorphous inorganic materials, especially for the amorphous lithium and sodium ionic conductors, including model construction, structural property and dynamical property calculation. The amorphous materials are usually metastable, and their DFT energies are non-ground state, higher than the corresponding the most stable crystalline phase. Therefore, those computational methods of assessing the chemical and electrochemical stability based on DFT ground-state energies are not available for the amorphous materials[64-65, 70], not summarized in this review.

### 2.1 Model construction

Model construction is the key for amorphous model, similar to sample preparation in the glass experimental research. Before model constructing, the size of the amorphous model must be carefully checked, which should be large enough to be realistic, and the computational resources will afford it. Since amorphous materials are disordered systems, each amorphous model is of limited size, and with specific composition and chemical components, it requires at least several samples for a targeted amorphous material. The results from reliable simulations, verified by the available experimental data, can lead to valid conclusions.

2.1.1  Initial amorphous structure

The formation of real amorphous materials comes from randomness of density fluctuations in the liquid (Brownian motions). However, the construction of a random packing of spheres is defined statistically over an ensemble, and therefore, it is a pseudo-random process. Random packing spheres is characterized by a simple algorithm in Eq. (1)[71]:

$$D + X \qquad \text{Eq. (1)}$$

Where, $D$ represents a sphere of diameter, and $X$ is the cut-off distance of all point-to-point (center of sphere). Amorphous structure models by random packing spheres can be built by the Packmol code[72]. Packmol is a command-line package for building the initial configurations with arbitrary chemical components of an amorphous material for MD simulations, which can generate a periodic amorphous structure starting with randomly packing N atoms, molecules, and ion groups in a defined cubic box with rough density and keeping a safe pairwise distance. Second, the continuous random network (CRN) models with perfectly coordinated atoms can be used for constructing covalent glasses and amorphous materials[73]. The amorphous structures based on the CRN model exhibit no coordination defect or void[74]. Third, Amorphous Cell module[75] based on the Monte Carlo method implemented in the Materials Studio software allows modeling the 3D periodic disordered atomic, molecular and polymer structures. Lastly, if the crystalline structures with the same chemical composition as the target amorphous materials are available, the amorphous structures also can be built from the fusing crystalline structures by MD simulations at high temperature (more than melt point)[76].

2.1.2 Melt-quench molecular dynamics method

In order to obtain the reasonable amorphous model of a target material, the melt-quench MD technique[76] can be utilized, which is the common method for modeling amorphous materials. In the melt-quench MD method, the choice of MD simulation is mainly categorized into two types[77-78]: the classical MD simulation based on the Newtonian mechanics, and the *ab*-initio molecular dynamics (AIMD) simulation based on the quantum mechanics. Due to their ab-initio level of accuracy and chemical versatility, AIMD simulations provide more accurate quantifications of the inter-atomic interaction than classical MD simulations. If the inter-atomic potential function and parameters for a target material are available, the classical MD simulation with large time-space scale can be used, and the total atom number and simulation time for the melt-quench MD can be more than ~1000 atoms and ~1 ns, respectively (Table 1). Otherwise, AIMD simulations have to be employed for the melt-quench MD procedure, which can only treat small system sizes and whose total atom number and simulation time are usually limited to within ~200 atoms and ~100 ps, respectively (Table 1). Moreover, very recently, an efficient and robust on-the-fly machine learning-driven AIMD method[79-81] is developed and integrated into the electronic-structure code, such as the Vienna Ab initio Simulation Package (VASP) software. Added by the automatically generated machine learning force fields from AIMD trajectories, the machine learning-driven AIMD simulations are not only accelerated by a factor of thousand compared with the pure ab initio calculations, but also provide good accuracy and transferability as same as the high-level ab-initio methods. Therefore, the machine learning-driven AIMD simulations can more effectively deal with the complex amorphous structures.

**Table 1.** Comparisons of the calculation methods for identifying amorphous structures.

|  | Simulation time length | Number of particles (typical values) | Accuracy | Computation cost |
| --- | --- | --- | --- | --- |
| RMC | - | ~10000 | Low | Low |
| Classical MD | ~1 ns | > 1000 | Medium | Medium |
| AIMD | ~100 ps | < 200 | High | High |
| MLIP-MD | ~1 ns | > 1000 | High | Medium |

Then, the heating and cooling procedure of the initial amorphous configurations with target chemical components can be performed by velocity rescaling on each MD step. The thermal procedure begins at RT (300 K), and the corresponding velocities are randomly initialized by the Maxwell-Boltzmann distribution[82]. The initial amorphous structure is then heated to a high temperature (more than melt point), which corresponds to a common heating rate of 1 K/fs. Note that selecting a high temperature for melting the amorphous structure should be careful. An appropriate high temperature can not only provide enough kinetic energies for atom diffusion to get the uniform amorphous structure, but also wouldn't distinctly tear the bonds of polyanion groups during the whole melt equilibrium process. In order to destroy any trace of the initial structure memory, MD runs at least 5 ps for fully melting the amorphous structure at this high temperature. Subsequently, the amorphous structure is cooled back to RT or 0 K with appropriate slow cooling rates, 0.02-0.05 K/fs[83]. The slow cooling rate should be much less than the heating rate, and the specific value of slow cooling rate should be determined from the tests of cooling rate dependent the quality of amorphous structures for the target materials. To prepare high-quality amorphous structures in a short time, the three-stage cooling procedure is recommended to be applied, namely "fast" + "slow" + "fast" cooling[76]. In the initial cooling procedure (stage-I), atoms in the box at high temperature are still in melting state and exhibit strong liquidlike behaviors, so this cooling interval can use the fast cooling rate. In the second temperature intervals (stage-II), being of great importance, this cooling procedure has to use the slow cooling rate to fully allow even improbable relaxation events to happen. In the last cooling procedure (stage-III), atoms in the box just exhibit vibrational displacements at the local minimum with almost no probability of a further structural relaxation. Consequently, the high cooling rate can be utilized again for this last cooling procedure to save computer resources. After reaching RT, the amorphous structures are evolved for the production stage to statistically average the structural and dynamical properties at RT, *e.g.*, the radial distribution function, structural factor, and power spectrum. Several independent "snapshots" are extracted from the production stage of 300 K MD simulations as the representative amorphous structures at RT.

The finished amorphous structures by the melt-quench technique can be further verified by the experimental results, such as radial distribution function (RDF) and powder X-ray diffraction (XRD) intensity. Recently, Jones et al. presented a structural descriptor of the glassy state by calculating the statistical averages within an ensemble framework[84]. In this method, the two quantities are used, the RDF, which captures the local order of the glassy state, and the XRD intensity, which captures the long-range order of the glassy state.

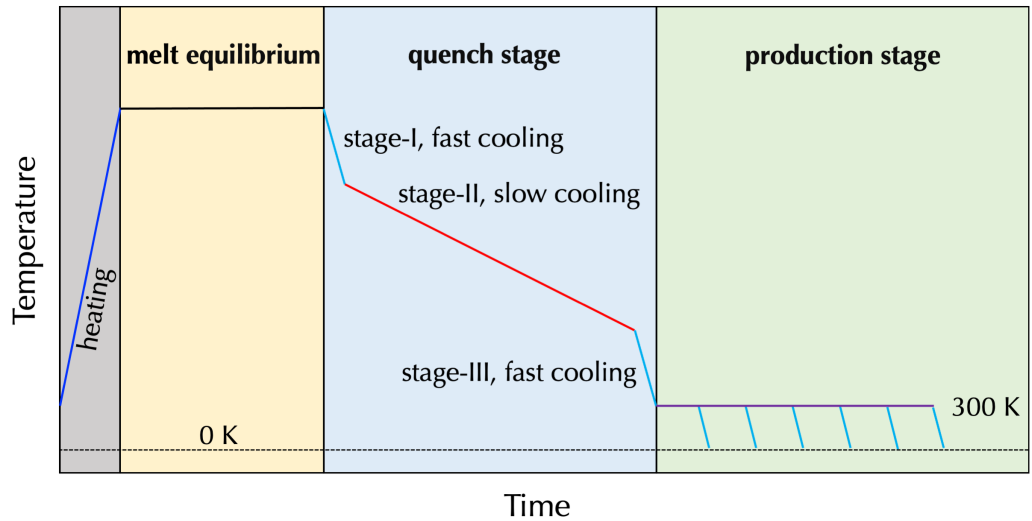

**Figure 1**. The scheme of melt-quench MD technique for modeling amorphous materials.

2.1.3 Reverse Monte Carlo simulation

Besides the melt-quench MD simulations, the reverse Monte Carlo (RMC) simulation is another useful tool for modelling the structures of glassy materials and further refined the MD quenched structure. It utilizes the standard Metropolis-Hastings algorithm[85] to address the inverse problem whereby the atomic configurations are continually adjusted until their properties have the greatest consistency with experimental data, such as the experimentally measured structure factor and radial distribution function[86]. Note that no input potential is required for RMC simulations, but they rely on the available experimental data to reproduce the atom-based structural models. RMC simulations can take account of the appropriate use of some constraints, such as coordination number and polyhedron connectivity.

To make a meaningful comparison between the computed and experiment determined PDF data, the computed PDFs should be broadened by convoluting the computed PDF with a normalized Gaussian distribution[87]. The level of agreement between the computed and experimental PDFs can be measured by the factor proposed by Wright[88-89]:

$$R_\chi^2 = \sum_i [g^{\exp}(r_i) - g^{\text{com}}(r_i)]^2 / \left(\sum_i g^{\exp}(r_i)\right)^2 \qquad \text{Eq. (2)}$$

A RMC simulation usually includes the following typical steps: (i) starting from a random initial configuration, PDF and the Wright's coefficient $R_\chi^{\text{old}}$ (Eq. (2)) of the simulated structure are calculated. (ii) a displacement is applied to a randomly selected atom, that is Monte Carlo move. Atomic displacements and directions should be randomly chosen, and they usually distribute between 0 and 0.2 Å. (iii) PDF and the Wright's coefficient $R_\chi^{\text{new}}$ of the new configuration are renewedly calculated. (iv) following the Metropolis algorithm[85], if $R_\chi^{\text{new}} \leq R_\chi^{\text{old}}$, the new configuration with a random atomic displacement applied can be accepted. Otherwise, that atomic displacement is accepted with the following probability (Eq. (3)):

$$P = \exp\left[-\frac{R_\chi^{\text{new}2} - R_\chi^{\text{old}2}}{T_\chi}\right] \qquad \text{Eq. (3)}$$

where $T_\chi$ is a unitless constant, controlling the acceptance probability of the Monte Carlo move. Now, RMC simulations are implemented in fullrmc[90], RMCProfile[91], RMC++[92], HRMC[93] software, etc.

## 2.1.4 Density determination

Especially, the calculated dynamical properties are quite sensitive to density or lattice volume[94]. Thus, density determination of the amorphous structure is very significant. If the experimental values of density for the target materials are available and firm, we can use them directly for MD simulations within the canonical ensemble (NVT). The second method is performing direct MD simulations at the target temperature within the Isothermal–isobaric (NPT) ensemble, in which the cell shape and volume of the amorphous structure model are variable during the whole MD running process[83, 95], and subsequently obtaining the average value of density. Both the classical MD and AIMD simulations are available for the direct MD simulation with NPT ensemble. AIMD simulations with variable-cells are usually performed by the Car–Parrinello molecule dynamics (CPMD) calculation[96], as implemented in the Quantum ESPRESSO[97] and CP2K[98] program packages. The third method is averaging the cell densities of a series of DFT optimized amorphous structures[99], which are quenched to 0 K in advance. Several independent "snapshots" extracted from the production stage of 300 K MD simulations are further quenched to 0 K by the high cooling rate, subsequently the thoroughly quenched structures (0 K) are fully optimized with both lattice constants and atomic coordinates to be relaxed to the ground state. After obtaining the correct density of the DFT ground state structure, the cells of the amorphous structure at other high temperatures are scaled to new volumes for MD simulations within the NVT ensemble.

2.1.5   Phase stability

Usually, the phase stability of a crystalline material can be theoretically evaluated by the corresponding compositional phase diagram[64-65, 70] , constructed by the DFT energy difference of this crystalline material and its compositional compounds from the Materials Project open database[100]. Whereas the amorphous materials are usually metastable, and their DFT energies are non-ground state, higher than the corresponding the most stable crystalline phase. Similarly, we can use the relative total energy difference between the amorphous and crystalline phase from the certain temperature MD simulations to access the phase stability of the amorphous material at this temperature. In MD simulations, the total energy of a model system is the sum of the potential energy and temperature dependent kinetic energy. If the relative total energy difference at temperature of $T$ is close to and even less than the value of ~$1.5k_BT$ ($k_B$ is the Boltzmann constant), the experimental synthesis at this temperature of this amorphous material by rapid quenching or energetic ball-milling is thermodynamically feasible. It should be noted that the contribution of the configuration entropy is not considered in MD simulations. Yet it is certain that the higher configuration entropies of amorphous structure compared to crystalline will further lower the Gibbs energy of the amorphous material, enhancing its phase stability to some extent.

**2.2   Structural properties**

The dominant features of short-range order and long-range disorder in the amorphous structures can be confirmed by calculating the radial distribution function and structural factors compared to the experimental data. In addition, the bond length, bond angle, coordination number and Voronoi index analysis are also done by computation for the structural analysis of the amorphous materials.

2.2.1 Radial distribution function

The RDF, $g(r)$, describes the probability of finding a particle at a distance of $r$ away from a given reference particle, relative to that for an ideal gas. The general algorithm involves determining how many particles are within a distance of $r$ and $r+dr$ away from a central particle at $r$. The partial RDF can be evaluated by Eq. (4) [71]:

$$g(r) = n(dr)/4\rho\pi r^2 dr \qquad \text{Eq. (4)}$$

where, $n(dr)$ is a function that computes the number of particles within a shell of thickness $dr$, $\rho$ is the numerical density of particles.

Amorphous materials follow the hard-sphere model of repulsion, indicating that there is zero density ($g(r)=0$) when atoms closely overlap each other. The first coordination sphere for an amorphous structure occurs at $r'$, where $r'$ is the maximum radius for $g(r)=0$. At large values of $r$, the atoms become independent of each other, and the system lose all of their long-range structures, making the distribution return to the average number density of bulk ($g(r)=1$). The first peak is the sharpest, indicating the presences of short-range order for the first coordination sphere in the amorphous structure. Subsequent peaks will occur roughly but be much smaller than the first peak. Second shell neighbors corresponding to the second peak, third shell neighbors and so on. Amorphous structures are more loosely packed than the crystalline phases, and therefore do not have exact intervals. The absence of long-range order in amorphous materials turns out to have significant effects on their vibrational, thermal and electronic properties.

2.2.2 Bond length and bond angle distribution

Usually, the radius of the first minimum ($r_{min}$) of $g(r)$ of an atom pair is regarded as the bond length of this atom pair. Bond angle is calculated by measuring the angle, $\theta$, between two vectors, originating from a reference atom and connecting two of its neighbors. The neighbor list includes all the atoms within a cutoff radius of the average bond length. The bond angle distribution function $f(\theta)$ measures the probability that the directions from a central atom to two of its neighbors form an angle $\theta$. $f(\theta)$ is essential a radial average over the triplet correlation function $g_3(r_1,r_2,\theta)$ over the nearest neighbor coordination shell[101], and the clear stepwise derivations of the triplet correlation function can be found in the previous studies[102-103]. Thus, $f(\theta)$ of the A-B-C angles can be calculated by Eq. (5):

$$f(\theta) = 16\pi^2 \int_0^{R_1} \int_0^{R_2} r_1^2 r_2^2 g(r_1)g(r_2)g_3(r_1,r_2,\theta)dr_1 dr_2 \qquad \text{Eq. (5)}$$

where $g(r_1)$ and $g(r_2)$ are RDFs of the B-A and B-C atom pairs, respectively. $R_1$ and $R_2$ are the bond lengths of B-A and B-C bonds, respectively, and are also equal to the first minimum radius of RDFs of the B-A and B-C atom pairs, respectively.

2.2.3 Coordination number

The coordination number (CN) indicates how many atoms are found in the range of each coordination sphere. A more general method for assigning spheres into positions relative to a central sphere is based on the radial distance from the center. Integrating $g(r)$ in spherical coordinates to the first minimum of the RDF of an atom pair will give the CN value of a central atom, which is given by Eq. (6)[104]:

$$\text{CN} = 4p\pi \int_0^{r_{min}} g(r) r^2 dr \qquad \text{Eq. (6)}$$

where $r_{min}$ is the radius of the first minimum of RDF of this atom pair.

On the other hand, CN also can be directly obtained from the coordination environment analysis based on the Voronoi tessellation approach[105-106] for determining the neighbors of a given atom, in which two parameters of distance and angle cut-offs should be carefully set. Automatically identifying the coordination environments of atoms in any material can be performed by the ChemEnv[107] module implemented in the open-source Pymatgen[100] code.

2.2.4 Structural factor

In contrast with crystals, amorphous structures are isotropic and have no long-range order, so their structure factors of S(q) only depend on the absolute magnitude of the scattering vector **q** and exhibit no sharp peaks. From MD simulations, the partial structure factors can be calculated by the Fourier transforming of the corresponding RDF of an atom pair, as shown in Eq. (7)[102-103]:

$$S_{\alpha\beta}(q) = 1 + 4p\pi \int_0^{r_{max}} w(r)(g_{\alpha\beta}(r) - 1)r^2 \frac{\sin(qr)}{qr} dr \qquad \text{Eq. (7)}$$

where $r_{max}$ is equal to half of the box length, and $w(r)$ is the weighting function applied to regulate the Fourier transforming, whose functional form can be referred from the previous work[103]. The total structure factor can be obtained by the weighted sum of all partial structure factors.

In addition, we can use another approach to directly calculate S(q) from the MD trajectories. If the position vectors of an atom *i* of specie $\alpha$ is labeled as $r_i$, then the Faber-Ziman partial structure factor, $S_{\alpha\beta}(q)$, would be defined as Eq. (8)[102]:

$$S_{\alpha\beta}(q) = 1 + \frac{N}{N_\alpha N_\beta} \left\langle \sum_i^{N_\alpha} \sum_{j \neq i}^{N_\beta} \exp\left(iq(r_i - r_j)\right) \right\rangle \qquad \text{Eq. (8)}$$

where $\langle \cdot \rangle$ stands for the statistical ensemble average. $N_\alpha$ and $N_\beta$ are the atom numbers of specie $\alpha$ and $\beta$, respectively, and *N* is the total number of all atoms. The wave vector *q* is ($n_x$, $n_y$, $n_z$)$2\pi/L$, where $n_i$ are integers and *L* is the length of simulation box. This method is very effective especially in the low *q* limit where the Fourier transforming method may fail to produce good S(q).

The total structure factors, obtained from the summation of $S_{\alpha\beta}(q)$ by multiplying appropriate coefficients, including the coherent bound neutron scattering lengths and X-ray scattering coefficient[78, 108], can be directly compared to the the structural factor data determined by the high-energy X-ray diffraction and neutron diffraction experiments, demonstrating the accuracy of MD simulations.

2.2.5 Voronoi tessellation analysis

For checking the micro-structures existent in the studied amorphous materials, we can carry out the Voronoi tessellation analysis[105], also called Voronoi diagram analysis, and the simulated atomic space is divided into some Voronoi polyhedra with the periodic boundary condition. Each individual Voronoi polyhedra is the basic unit of the structural space of the amorphous material, which is similar to the unitcell of the crystalline material. The Voronoi polyhedra possess unambiguously topological indexes, topologically reflecting the geometry distributions and CN of a local atomic structure, and these basic attributes in topology can be used for describing the local structural characteristics of the amorphous materials. In a topological index, the first number is the number of those faces surrounding the central atom, containing three edges; the second number is the number of those faces composed of four edges surrounding the central atom, and so on. For example, the simple *fcc* structures process the Voronoi polyhedral index of [0 12 0 0]. That is, there are 12 tetragonal (the nearest neighbor atoms) planes in *fcc* structures without any other component planes, and the corresponding CN is 14. While the polyhedral index of *bcc* structures is [0 6 0 8], so their Voronoi polyhedra are composed of 6 pentagons (the second-nearest neighbor atoms) and 8 hexagon (the nearest neighbor atoms) planes, with a CN of 14. Now, computing the Voronoi tessellation of an amorphous structure can be performed by the Materials Studio, MatLab, Mathematica, and Maple softwares.

## 2.3 Dynamical properties

Dynamical properties are the keys for SSE materials. MD simulations sample the statistical contributions of all ion diffusion events and quantify the diffusional properties, such as ionic diffusivity, ionic conductivity and activation energy barrier[109]. In addition, MD simulations played unique roles in identifying the correlated diffusion mechanisms in some lithium (sodium) superionic conductors[110-111], such as $Li_7La_3Zr_2O_{12}$[112-113], $Li_7P_3S_{11}$[17] and $Na_3Zr_2Si_2PO_{12}$[114], rather than the isolated ion hopping often assumed in some previous nudged elastic band (NEB) calculations[115-116].

### 2.3.1 Mean squared displacement

The diffusional properties can be calculated from the MD trajectories of particles with positions of $r_i(t)$, and the displacement $\Delta r_i$ of *i*-th particle from time $t_1$ to $t_2$ can be calculated by Eq. (9):

$$\Delta r_i(\Delta t) = r_i(t_2) - r_i(t_1), \text{where } \Delta t = t_2 - t_1 \qquad \text{Eq. (9)}$$

The total squared displacements are obtained by summing the squared displacements of all $N$ mobile particles over a time interval $\Delta t$, $\sum_{i=1}^{N}(|\Delta r_i(\Delta t)|^2)$. Over a total MD simulation time duration of $t_{total}$, there are many time intervals of $N_{\Delta t}$ with the same duration of $\Delta t$ ($\Delta t < t_{total}$) at different starting time of $t$. Due to the particle's displacements over $\Delta t$ reflecting the mobility of particles, the total mean squared displacements (TMSDs) of all $N$ mobile particles over a time interval of $\Delta t$ can be obtained by calculating the statistical ensemble average of the squared displacements over a total of $N_{\Delta t}$ time intervals with the same duration of $\Delta t$ [78]:

$$\text{TMSD}(\Delta t) = \sum_{i=1}^{N}\langle|r_i(\Delta t) - r_i(0)|^2\rangle = \sum_{N=1}^{N}\frac{1}{N_{\Delta t}}\sum_{t=0}^{t_{total}-\Delta t}|r_i(t+\Delta t) - r_i(t)|^2 \qquad \text{Eq. (10)}$$

where $\langle \cdot \rangle$ stands for the statistical ensemble average. This statistical ensemble average over different time intervals of $N_{\Delta t}$ offers the statistical analysis of sufficient diffusional events to obtain accurate diffusional properties. To get the diffusivity of the mobile-particle species, the MSD over time interval of $\Delta t$ is calculated by averaging TMSDs to each mobile particle:

$$\text{MSD}(\Delta t) = \text{TMSDs}(\Delta t)/N \qquad \text{Eq. (11)}$$

where, $N$ is the number of particles regarded as the mobile carriers contributing to diffusion.

2.3.2 Diffusivity and ionic conductivity

If MD simulations contain sufficient diffusional event, the dependence of MSD over time intervals of $\Delta t$ would follow a linear relationship. According to the Einstein relation, the self (tracer)-diffusion coefficient $D$ of a specie can be calculated from the slope of MSD curve as a function of time intervals $\Delta t$ by Eq. (12)[117]:

$$D = \frac{\text{MSD}(\Delta t)}{2d\Delta t} \quad \text{Eq. (12)}$$

where, $d=3$ is the diffusion dimension of a particle in the simulated system. The linear fitting of MSD vs. $\Delta t$ to the Einstein relation should only be performed on the linear region corresponding to good diffusional displacements, excluding the initial region with ballistic motion displacement and the final region with a small number of time time interval $N_{\Delta t}$. Therefore, to achieve small error bounds of the fitted D and get accurate diffusivity, MD simulations should be long enough for capturing a large number of diffusion events. Additionally, MSD per mobile particle should be larger than a few times of $a^2$ to pick out the ballistic region[109], where $a$ is the distance between two neighboring occupation sites of mobile particle.

This calculated self (tracer)-diffusion coefficient $D$ is the tracer diffusivity of the mobile-particle specie, and is an intrinsic property of the mobile-ion specie under certain conditions. With the self (tracer)-diffusion coefficient, the ionic conductivity $\sigma$ is calculated according to the Nernst–Einstein relation by Eq. (13)[118]:

$$\sigma = \frac{Nq^2}{Vk_BT}D \quad \text{Eq. (13)}$$

where *V* is the total volume of the simulated model system, *q* is the ionic charge of the mobile specie regarded as carrier contributing to conductivity, *T* is the absolute temperature, and $k_B$ is the Boltzmann constant. By combining Eqs. (11–13), the ionic conductivity is calculated by Eq. (14):

$$\sigma = \frac{q^2}{Vk_BT}\frac{\text{TMSD}(\Delta t)}{\Delta t} \qquad \text{Eq. (14)}$$

Performing a series of MD simulations at different temperatures obtains the Arrhenius relations of the log of diffusivity *D* as a function of 1/*T* (Eq. (15)), and which can extrapolate the prefactor, overall activation energy $E_a$, diffusivity *D* and conductivity *σ* at *RT*, which is similar to experimental measurements.

$$D = D_0 \exp(-E_a/k_BT) \qquad \text{Eq. (15)}$$

where, $D_0$ is known as the pre-exponential factor.

2.3.3 Correlated diffusion

The Nernst–Einstein relation (Eq. (13)) from self (tracer)-diffusion coefficient *D* assumes dilute, isolated mobile-ion carriers in the materials systems. The failure of the Nernst-Einstein relation is noticed in LGPS SSE material, and its application leads to an underestimation of the value of ionic conductivity due to the correlated lithium ion diffusion[111]. Considering the effect of ionic correlation, the ionic diffusivity is calculated from MSD of net ionic motion by Eq. (16)[119-120]:

$$D_\sigma = \frac{1}{2d\Delta t} \left\langle \left| \frac{1}{N}\sum_{i=1}^{N} r_i(t+\Delta t) - \frac{1}{N}\sum_{i=1}^{N} r_i(t) \right|^2 \right\rangle \quad \text{Eq. (16)}$$

A measured tracer self (tracer)-diffusion coefficient $D$ can be smaller than $D_\sigma$. A correlation factor, $f = ND_\sigma/D$, is often used to quantify the correlation of particles[121], which is described by the ratio of the random-walk diffusion coefficient $D_\sigma$ of all N particles against the tracer self (tracer)-diffusivity $D$ calculated from Eq. (12).

In addition, the correlated or cooperative motion can be quantified by computing the van Hove correlation function. The van Hove correlation can be divided into the self-part $G_s$ and the distinct-part $G_d$, as following Eqs. (17-18)[78, 122-123]:

$$G_s(r,t) = \frac{1}{4\pi r^2 N} \left\langle \sum_{i \neq j}^{N} \delta(r - |r_i(t_0) - r_j(t+t_0)|) \right\rangle \quad \text{Eq. (17)}$$

$$G_d(r,t) = \frac{1}{4\pi r^2 \rho N} \left\langle \sum_{i \neq j}^{N} \delta(r - |r_i(t_0) - r_j(t+t_0)|) \right\rangle \quad \text{Eq. (18)}$$

where $\delta(x)$ is the one-dimensional Dirac delta function, and the average number density $\rho$ normalizes $G_d$ to 1 when $r \gg 1$. $G_s(r, t)$ defines the probability of a particle diffusing away from its initial site by a displacement of $r$ over time $t$, whereas $G_d(r, t)$ characterizes the real-space radial distribution function of $N$-1 particles over time $t$ with respect to the initial reference particle in the lattice. When $t=0$, $G_d(r, 0)$ is reduced to the static RDF. When $r=0$, $G_d(0, t)$ can show the correlation time of one vacant particle site being occupied by another particle. Now, the van Hove function analyses based on MD trajectories are available in the open-source pymatgen-diffusion package.

2.3.4 Lattice dynamics

Some recent works on superionic conductors show the influence of lattice dynamics on ionic transport[124-127]. A softer lattice would lower ionic migration enthalpy. Indeed, as ionic migration is a thermally activated process, it is expected that ionic diffusivity should be related to the Debye frequency and vibration strength of lattice. The lattice dynamics information, such as amplitude of vibrations, attempt frequency, jump rates and vibration frequency spectrum (vibrational density of states, VDOS) of mobile ions, can be obtained from MD simulations, which fully take into consideration anharmonicity and temperature effects[128-129]. The VDOS of mobile ions can be calculated by the Fourier transform of the corresponding velocity-velocity autocorrelation function from MD simulations, as shown in the following Eq. (19):

$$\text{VDOS}(\omega) = \frac{1}{N}\sum_i^N \langle v_i(t) \cdot v_i(0) \rangle e^{iwt} dt \qquad \text{Eq. (19)}$$

where $v_i(t)$ represent the velocity of ion *i* at time *t*. Furthermore, based on known VDOS, it is straightforward to calculate the corresponding phonon band center or average lattice vibration frequency $\bar{\omega}$, defined as Eq. (20)[124]:

$$\bar{\omega} = \frac{\int_0^\infty \omega * \text{VDOS}(\omega) d\omega}{\int_0^\infty \text{VDOS}(\omega) d\omega} \qquad \text{Eq. (20)}$$

2.3.5 Coupled cation−anion dynamics

The coupled cation–anion dynamics (paddle-wheel effect) contributions to ionic mobility can be captured by analyzing spatial, temporal, vibrational, and energetic correlations[95, 130-131]. The rotations/reorientations of chalcogenides X anion can be evaluated by the normalized 2D probability density distribution $\rho^{2D}_{(\theta,\varphi)}$ as a function of $\theta$ and $\varphi$ for the chalcogenides X ligands in the PnX$_n$ polyanions from MD simulations. Here, $\theta$ corresponds to the angle between a selected Pn-X bond and the *z* axis, and $\varphi$ donates the angle between the projection of the Pn-X vector in the *xy* plane and the *x* axis. As $\rho^{2D}_{(\theta,\varphi)}$ is known, the Helmholtz free energy surface *H* of the chalcogenides X ligands can be further calculated as Eq. (21)[132]:

$$H(\theta, \varphi) = -k_B T \ln\left[\rho^{2D}_{(\theta,\varphi)}\right] \qquad \text{Eq. (21)}$$

By the obtained Helmholtz free energy surface, the activation energy barrier of PnX$_n$ polyanion rotations/reorientations can be defined by the energy difference between the local minimum and the saddle point.

To quantify the paddle-wheel effect (correlation) between the mobile-ion (*e.g.*, Li$^+$, Na$^+$) translational diffusion and the rotation of the chalcogenide ligands (X) in the PnX$_n$ polyanions quantitatively, the 2D-joint probability distribution $\rho^{2D}_{(\theta,r)}$ as a function of the angle $\theta$ of Pn-X–Li bond and the distance *r* between X and Li would be employed. The square root of $\rho^{2D}_{(\theta,r)}$ is defined as Eq. (22)[131]:

$$f(\theta, r) = \sqrt{\rho^{2D}_{(\theta,r)}} \qquad \text{Eq. (22)}$$

and its uncorrelated counterpart of the one-dimensional (1D) probability distribution as a function of the angle $\theta$ of Pn-X–Li bond and the distance $r$ between X and Li, respectively:

$$g(\theta, r) = \sqrt{\rho_\theta^{1D} \rho_r^{1D}} \qquad \text{Eq. (23)}$$

If there is a correlation between $\theta$ and $r$, the following quantity, $\chi$, represents the correlation coefficient:

$$\int d\theta dr [f(\theta, r) - g(\theta, r)]^2 = \chi^2 \quad \text{Eq. (24)}$$

$\chi = 0$ means that there is no correlation between $\theta$ and $r$. Intermediate values of $\chi$ indicate some correlation between $\theta$ and $r$.

Moreover, the cation translational diffusion and anion rotation needs to occur at the same timescale (with same frequency) to ensure the coupled cation–anion dynamics. The power spectrum of both cations and anions can be obtained by the Fourier transform of the corresponding linear and angular velocity autocorrelation function from MD simulations, respectively. The low frequency overlapping of power spectrum is the evident of the coupled motion and probable momentum transfer between cation diffusion and anion rotation. The linear power spectrum of cation translational diffusion is calculated from the Fourier transform of the averaged linear velocity autocorrelation function $\langle \sum \hat{v}_{cation}(t) \cdot \hat{v}_{cation}(t + \Delta t) \rangle$, where $\hat{v}_{cation}(t)$ is the linear velocity of cation translational diffusion. The angular power spectrum for anion rotation is obtained via the Fourier transform of the averaged angular velocity autocorrelation function $\langle \sum \hat{\omega}_{\text{anion}}(t) \cdot \hat{\omega}_{\text{anion}}(t + \Delta t) \rangle$, in which $\hat{\omega}_{\text{anion}}(t)$ is the angular velocity of the anion X ligand in PnX$_4$ polyanion groups at time $t$, which is calculated via Eq. (25):

$$\hat{\omega}_{\text{anion}} = \frac{\hat{r}_{\text{anion}} \times \hat{v}_{\text{anion}}}{r_{\text{anion}}^2} \quad \text{Eq. (25)}$$

Here, $\hat{r}_{\text{anion}}$ and $\hat{v}_{\text{anion}}$ are the position vector and velocity for each X anion relative to the central of mass of PnX$_4$ polyanion.

## 2.4 Workflow of high-throughput AIMD calculation of amorphous materials

MPmorph is an open infrastructure for running and analyzing AIMD calculations of the amorphous materials (https://github.com/materialsproject/mpmorph)[133]. MPmorph provides the powerful workflows for massively generating amorphous configurations with molar volumes 20% larger than those of the ground-state crystal structures at the same composition in the Materials Project (MP) database[134]. MPmorph affords the automatic workflows of AIMD simulations running with VASP (Vienna *ab* initio simulation package) for the melt-quench process, and provides some powerful tools for statistical analysis of the structural and dynamical properties, including RDF, CN, Voronoi tessellation analysis, polyhedron connectivity, thermodynamic quantities and diffusion coefficients.

## 2.5 Machine learning for material research

With the rapid growth of materials databases, machine learning techniques have successfully made many breakthroughs in the field of energy storage and conversion materials[135], such as catalysts[136-138], photovoltaics[139-141], and battery materials[142-144]. Expect for the energy related materials, so far, machine learning models and predictions have been utilized for accelerating the invention and research processes of amorphous and disordered materials of metallic glasses[145-146], such as Co-V-Zr[147] and Cu-Zr[148]. As a whole, there are three main modes of applying the machine learning method to glassy material research, including data-driven machine learning, machine learning interatomic potential, and machine learning-driven AIMD simulation.

For the data-driven machine learning, sufficient structure, dynamics, mechanical properties and other data from the high-throughput experiments and calculations are analyzed to establish the relationship between structure and performance, and build the descriptor-property models to give theoretical predictions. Note that the biggest challenge in applying machine learning to the field of amorphous materials is the establishment of a relevant database containing sufficient data. When the current database has not been established or has not been completely established, there are some ways to apply machine learning method when the amount of data is insufficient. On the one hand, some machine learning methods for processing small sample data have been developed, such as Naive Bayes algorithm, decision tree, and other linear models. Generally, the simpler the machine learning algorithm is, the better the processing small sample data will be. Because the sample data requires a low-complexity model to avoid overfitting. On the other hand, after preprocessing all the data using certain methods, the accuracy of model prediction can be improved. Analyzing the data set and machine learning prediction ability, it is found that small sample data does not directly affect the accuracy of model[149], but the degree of freedom of model is used as an intermediary, resulting in a correlation between accuracy and freedom. However, to increase the accuracy by adding additional experimental data, a large amount of experimental cost needs, and the cost is not directly proportional to the improvement of accuracy. Therefore, the key challenges of practicing machine learning in feature modeling are improve the accuracy of model by some methods without the need for higher model degrees of freedom.

Machine learning interatomic potentials are fitting the interatomic potentials for classical MD simulations based on the machine learning method, called as the machine learning interatomic potential molecule dynamics (MLIP-MD) simulation. Machine learning potentials can often give good predictions of alloy properties, such as energy and force with near-DFT accuracy but are orders of magnitude faster than the pure first-principle calculations[150-153]. At this stage, the research on the potential function fitting is mainly based on the high-dimensional neural network potential (NNP)[154], the Gaussian approximation potential (GAP)[153], the spectral neighbor analysis potential (SNAP)[155], and moment tensor potentials (MTP)[156]. However, the applications of machine learning interatomic potential approaches are still limited to some simple metals, alloys and binary compounds. These difficulties mainly come from the force field generation process, including complex train data selection and parameter optimization. In order to well train force fields, we need carefully selected reference datasets of structures, energies and forces from the first-principle calculations[157-158].

Wang et al. utilized MLIP-MD simulations based on the moment tensor potentials to calculate lithium-ion diffusivities, significantly increasing the efficiency of the calculations by seven orders of magnitude compared to pure AIMD simulations[159]. Huang et al. developed an efficient protocol to automatically generate the neural network potentials for $Li_{10}GeP_2S_{12}$-type SSE materials, extending the simulation systems with large sizes (~1000 atoms) and carefully investigating the statistical errors and size effects[160]. Park et al. developed a graph neural network framework to directly predict atomic forces of $Li_7P_3S_{11}$ crystal and performed MD simulations to get Li diffusion coefficient, which is within 14% of that obtained directly from AIMD[161]. The combination of a genetic algorithm and a specialized artificial neural networks (ANNs) potential was used to highly speed up the sampling of amorphous materials[162-163]. Assisted by the ANN-potential, around 1000 first-principles calculations are sufficient for sampling all the low-energy atomic configurations in the entire amorphous $Li_xSi_y$ alloy phase space[162]. In addition, Qi et al. revealed that the lithium-ion diffusivities of $Li_{0.33}La_{0.56}TiO_3$, $Li_3YCl_6$ and $Li_7P_3S_{11}$ SSEs undergo the non-Arrhenius behaviors at relatively low temperatures by performing the low-temperature (less than 400K) and long-time MLIP-MD simulations with MTP, which well bridge the gaps between the predicted and experimentally measured ionic conductivities and activation energies[164]. However, so far, few machine learning interatomic potentials have been fitted for the amorphous materials[165], thus, the long-time and large spatial scale MLIP-MD simulations with machine learning fitting potentials for the glassy SSE materials are quite interesting and worthwhile thing to do.

Different from the machine learning interatomic potentials for classical MD simulation, the machine learning force field can be also integrated into an electronic-structure code, and it realizes automatic generation of on-the-fly machine learning force fields on the basis of Bayesian inference during MD simulations, where the first-principles calculations are only executed, called as the machine learning-driven AIMD simulations[81]. With the help of the on-the-fly machine learning force fields, machine learning-driven AIMD simulations are accelerated by a factor of thousand compared with the pure AIMD calculations. It is successfully utilized to study the liquid-solid phase transitions of a wide variety of materials, obtaining near-DFT level results, *e.g.*, the calculation of melting points of Al, Si, Ge, Sn and MgO[81]. Now, the machine learning-driven AIMD simulation method has been implemented in the latest version VASP code. It is conceivable that machine learning-driven AIMD method can more effectively deal with MD simulations of the complex glass structures.

## 3. Lithium sulfide-type glasses

This and next sections of reviewing on the computational investigations are mainly organized in the order of the publication time of their work. Early in 2006, Seshasayee et al. performed the classical MD simulations for the ternary $Li_2S$-$P_2S_5$-LiI glasses with utilization of the melt-quench technique[166]. Interestingly, there are two separated regions for the $0.375Li_2S$-$0.375P_2S_5$-$0.25LiI$ glass, including I-rich region and I-poor region, as illustrated in **Figure 2a**. The shortest Li–I distance observed here is close to 6Å, and Li atoms are only surrounded by sulfurs anions. No separated region is observed from the $Li_2S$-$P_2S_5$-LiI glasses with other components. The two separated I-rich and I-poor regions provide more pathways for easy movement of Li ion compared to other glasses with different compositions[166], consequently this glass shows higher lithium ionic conductivity at RT determined by both experiment and MD calculation. This specific phenomenon of I atom separation isn't due to the absence of Li-I interaction, and it may be an intrinsic characteristic of the $0.375Li_2S$-$0.375P_2S_5$-$0.25LiI$ glass with this given component, which needs the relevant experimental data to verify, e.g., elemental line scanning analysis for the rapidly quenched glassy structure by X-ray energy spectrum.

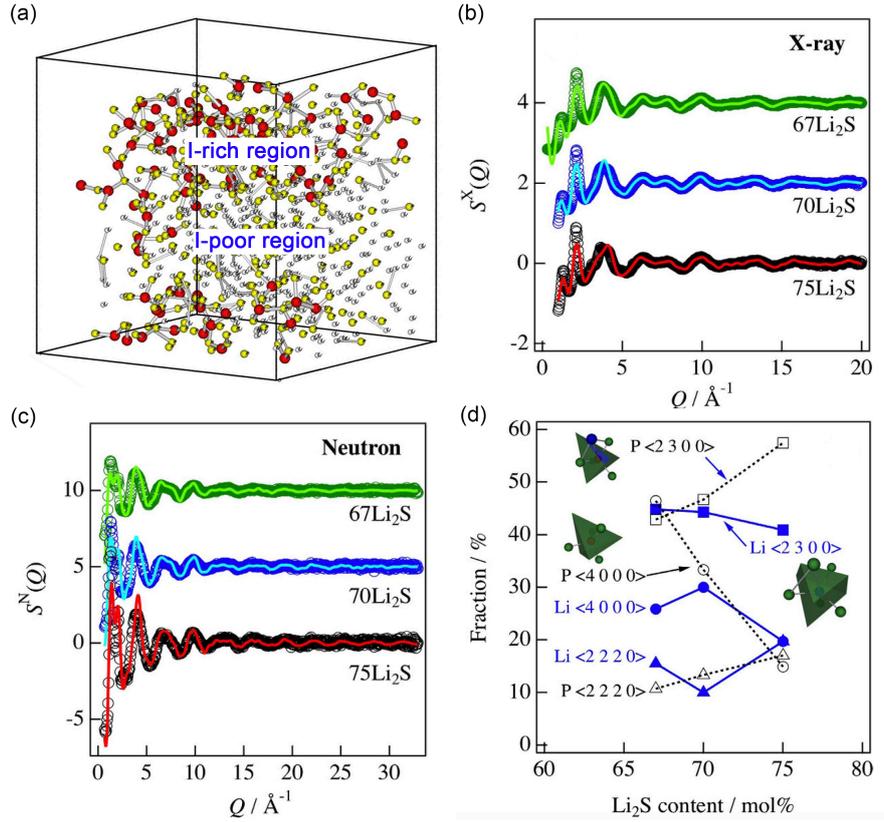

**Figure 2**. (a) Atomic configuration with two separated I-rich and I-poor regions of the $0.375Li_2S$-$0.375P_2S_5$-$0.25LiI$ glass at 300 K (Red, yellow and gray balls donate iodine, phosphorus and lithium atoms, respectively, and sulfur atoms are not shown for clarity). Total structural factors $S(q)$ at RT for $xLi_2S$-$(1-x)P_2S_5$ ($x=0.67$, 0.70 and 0.75) glasses derived from (b) X-ray and (c) neutron diffraction (circles, experimental data; lines, RMC model), (d) P-centered and Li-centered Voronoi polyhedra for $Li_2S$-$P_2S_5$ ($x=0.67$, 0.70 and 0.75) glasses derived from DFT/RMC simulations. Part (a) is reproduced with permission from reference[166], (Copyright 2006, Elsevier). Part (b-d) are reproduced with permission from reference[35], (Copyright 2016, Nature).

A series of the RMC simulations have been done to model the atomic configurations for $Li_2S$-$P_2S_5$ superionic glasses based on the experimental data of synchrotron radiation X-ray and neutron diffraction measurements, including $0.2Li_2S$-$0.8P_2S_5$[167], $0.4Li_2S$-$0.6P_2S_5$[167], $0.5Li_2S$-$0.5P_2S_5$[168], $0.6Li_2S$-$0.4P_2S_5$[167-168], $0.67Li_2S$-$0.33P_2S_5$[35], $0.7Li_2S$-$0.3P_2S_5$[35, 167, 169], and $0.75Li_2S$-$0.25P_2S_5$ glass[35]. These RMC simulations were performed in those boxes with ~5000 atoms, and three constraints were taken into consideration: (1) the $PS_4$ tetrahedron is composed of bridging S and non-bridging S; (2) the CN of P around bridging S must be 2; (3) the CN of P around non-bridging S must be 1. RMC modeling can well reproduce the plausible atomic configurations for $Li_2S$-$P_2S_5$ superionic glasses, providing excellent fitting between the experiment observed and calculated structural factor, $S(q)$, as illustrated in **Figure 2b** and **2c**. RMC simulations of $xLi_2S$-$(1-x)P_2S_5$ glasses reveal that for the lower $Li_2S$ content (x ≤ 40), $Li^+$ cations are distributed individually in the $PS_4$ tetrahedra networks, whereas for the higher $Li_2S$ content (x ≥ 60), the short-range $Li^+$-$Li^+$ correlations within 4.0 Å drastically increase. By the DFT/RMC simulations of $xLi_2S$-$(1-x)P_2S_5$ glasses[35], the Voronoi polyhedron statistics were calculated to obtain the local coordination environment of a central atom in detail (see **Figure 2d**). The atom-centered Voronoi polyhedra includes information about the coordination environment beyond the first coordination shell. The fractions of P-centered Voronoi polyhedra with index of [2 3 0 0] and [2 2 2 0] beyond the first coordination environment increase relative to that of the [4 0 0 0] P-centered Voronoi polyhedra ($PS_4$) with the increase of $Li_2S$ content. The simple Li-centered Voronoi polyhedra with index of [4 0 0 0] and [2 3 0 0] are dominant in $Li_2S$-$P_2S_5$ glass, suggesting that the free volume around $PS_x$ polyanions allows the $Li^+$ ion distributions within the first coordination shell to form $LiS_4$ and $LiS_6$ polyhedra even at a higher $Li_2S$ content.

RMC simulations promote the understanding of glass materials, but they only provide the structural information and no time-dependent dynamics properties can be available.

Beyond the RMC models, MD simulations based on solving the time-dependent Newton's motion equation not only reproduce the atom structures, but also provide the dynamics properties, such as atom diffusion and VDOS of a mobile atom. Therefore, MD simulations have been widely employed for studying the structural and dynamical properties of glassy SSE materials. A pioneering work was reported in 2016 by Kawamura et al.[83], and they successfully modeled the $Li_2S$-$P_2S_5$ glassy SSEs with amorphous structures by the melt-quench MD method with NPT ensemble, and investigated $Li^+$ ion diffusions by employing MD calculations based on density functional theory (DFT-MD), which can be also called as AIMD calculations. In their work, the initial structures of $xLi_2S$-$(100-x)P_2S_5$ (x = 0.67, 0.70, 0.75, and 0.80) glasses were built by packing the ionic units of $Li^+$, $PS_4^{3-}$, $P_2S_7^{4-}$, and $S^{2-}$, in which the covalent bonding between P and S in polyanions were considered to make the initial packed structures more closed to the reasonable ones. AIMD simulations well reproduced the amorphous structures of these metastable $Li_2S$-$P_2S_5$ glasses, which are comparable to the experimental data, including X-ray and neutron diffraction determined structure factors. During the AIMD simulations, the polyanion units, including tetrahedral $PS_4^{3-}$ and di-tetrahedral $P_2S_7^{4-}$, were well maintained and randomly distributed in the periodical boxes, as an example illustrated in **Figure 3a** (a geometrical snapshot in AIMD simulations at 300 K of the amorphous structure of $0.7Li_2S$-$0.3P_2S_5$), indicating the existence of these polyanion units is the intrinsic characteristics of each $Li_2S$-$P_2S_5$ glass. AIMD simulations clearly show the calculated mass densities of $Li_2S$-$P_2S_5$ glasses show a peak at x = 70 (see **Figure 3b**), which is not the same as the order of the molecular ratios of $Li_2S$. Note that only the 0.7Li$_2$S-0.3P$_2$S$_5$ glass synthesized by the mechanical milling method has a experimental density value of 1.93 g/cm$^3$ [35], and the AIMD calculated density of 1.89 g/cm$^3$ for the 0.7Li$_2$S-0.3P$_2$S$_5$ glass is very closed to the experimental result. Compared to the densities of the corresponding crystals determined both by experiments and geometrical optimization of DFT calculation, the densities of Li$_2$S-P$_2$S$_5$ glasses (1.78-1.89 g/cm$^3$) are lower than those of the corresponding crystals by 5-8%, which is much smaller than the relative density changes of other inorganic glasses[170].

Moreover, AIMD simulations uncovered the lithium ion transport properties of xLi$_2$S-(100-x)P$_2$S$_5$ (x = 0.67, 0.70, 0.75, and 0.80) glasses. The duration time (40 ps) of their AIMD simulations is much shorter than the typical time settings in the classical MD simulations, but the accurate atomic forces obtained by the DFT method is important to describe Li diffusion precisely. At 300 K, almost all (99.1%) lithium ions move within 2.5 Å, and most of lithium ions stay at their own sites and couldn't get the nearest neighbor sites, indicating the weak lithium ion transport in these Li$_2$S-P$_2$S$_5$ glass at room temperature. More lithium ions move over 4.5 Å from the initial sites to the second nearest neighbor sites at higher temperatures (9.5, 36.5, and 85.2% of lithium ions migrate over 4.5Å at 500, 667, and 1000K, respectively). The highest calculated ionic conductivity of the xLi$_2$S-(100-x)P$_2$S$_5$ glasses corresponds to a Li$_2$S concentration of x = 0.75, which is consistent with experimental data. Although the lithium ion number density of 0.8Li$_2$S-0.2P$_2$S$_5$ is the highest, the calculated ionic conductivity is the lowest, due to the isolated sulfur ions with greater negative charges only existing in the 0.8Li$_2$S-0.2P$_2$S$_5$ glass would significantly suppressing lithium ion migration. Therefore, there is balance between the lithium ion number density and diffusion coefficient to achieve high ionic conductivity for the Li$_2$S-P$_2$S$_5$ glasses with different Li$_2$S concentrations.

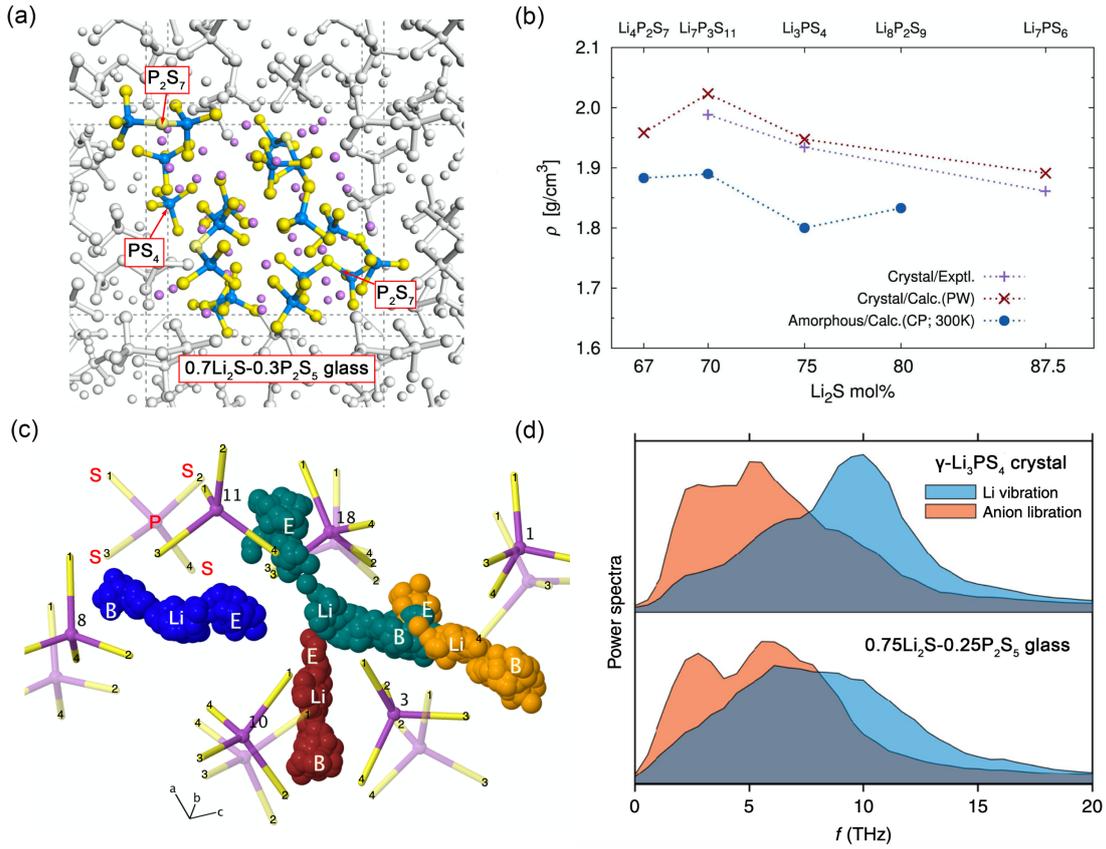

**Figure 3**. (a) A trajectory snapshot from 300K AIMD simulation of 0.7Li$_2$S-0.3P$_2$S$_5$ glass, the tetrahedral PS$_4^{3-}$ and di-tetrahedral P$_2$S$_7^{4-}$ coexist in the periodical box, purple, blue, and yellow balls donate Li, P, and S atoms, respectively; (b) AIMD simulations calculated mass densities, $\rho$, of the xLi$_2$S-(100-x)P$_2$S$_5$ (x = 0.67, 0.70, 0.75, and 0.80) glasses and comparisons with the corresponding crystal densities determined by experiments and DFT calculations; (c) the cation–anion coupled diffusion mechanism in 0.75Li$_2$S-0.25P$_2$S$_5$ glass at 300 K, and distinct colored spheres represent the translational diffusions of four different lithium ions over a 10 ps trajectory; the initial and final positions of lithium ions are labeled "B" and "E", respectively; (d) power spectra obtained from Fourier transform of the Li$^+$ velocity autocorrelation and angular velocity autocorrelation functions of PS$_4^{3-}$ rotation for crystalline γ-Li$_3$PS$_4$ and 0.75Li$_2$S-0.25P$_2$S$_5$ glass at 300 K. Parts (a-b) are reproduced with permission from reference[83], (Copyright 2016, Baba and Kawamura). Parts (c-d) are reproduced with permission from reference[95], (Copyright 2020,

Nature).

Sadowski et al. present a AIMD theoretical work of xLi$_2$S-(1-x)P$_2$S$_5$ ($x$ = 0.67, 0.70 and 0.75) glasses, including the sulfur-deficient compositions towards Li$_4$P$_2$S$_6$[171]. The thermodynamic stability of a glassy structure is well-connected with the species of P$_x$S$_y$ structural units, and it decreases from P$_2$S$_7^{4-}$ over PS$_4^{3-}$ to P$_2$S$_6^{4-}$ (**Figure 4a**). Calculated Arrhenius plot (**Figure 4b**) shows the similar lithium transport properties for all glass system regardless of the chemical compositions and underlying structural units, due to all P$_x$S$_y$ structural units provide very similar chemical environments for lithium. Moreover, in the xLi$_2$S-(1-x)P$_2$S$_5$ ($x$ = 0.67, 0.70 and 0.75) glasses, they observed the highly correlated and concerted lithium ion motion. In the slowly quenched glassy structures, the ''unusual'' structural units (**Figure 4c**) with partly low excess energies were formed, and they found lower quenching rates would improve the stability. Under the condition of local lithium deficiency (e.g., close to electrode interfaces), as shown in **Figure 4d**, there are some the intermolecular S–S dimers between neighboring P$_x$S$_y$ structural units due the oxidation effect of sulfur anions. These S–S dimers and the ''unusual'' P$_x$S$_y$ structural units would lower the electronic band gap, and promote the mixed electronic–ionic conduction, eventually initiating the degradation reactions of the sulfur-deficient Li$_2$S-P$_2$S$_5$ glasses.

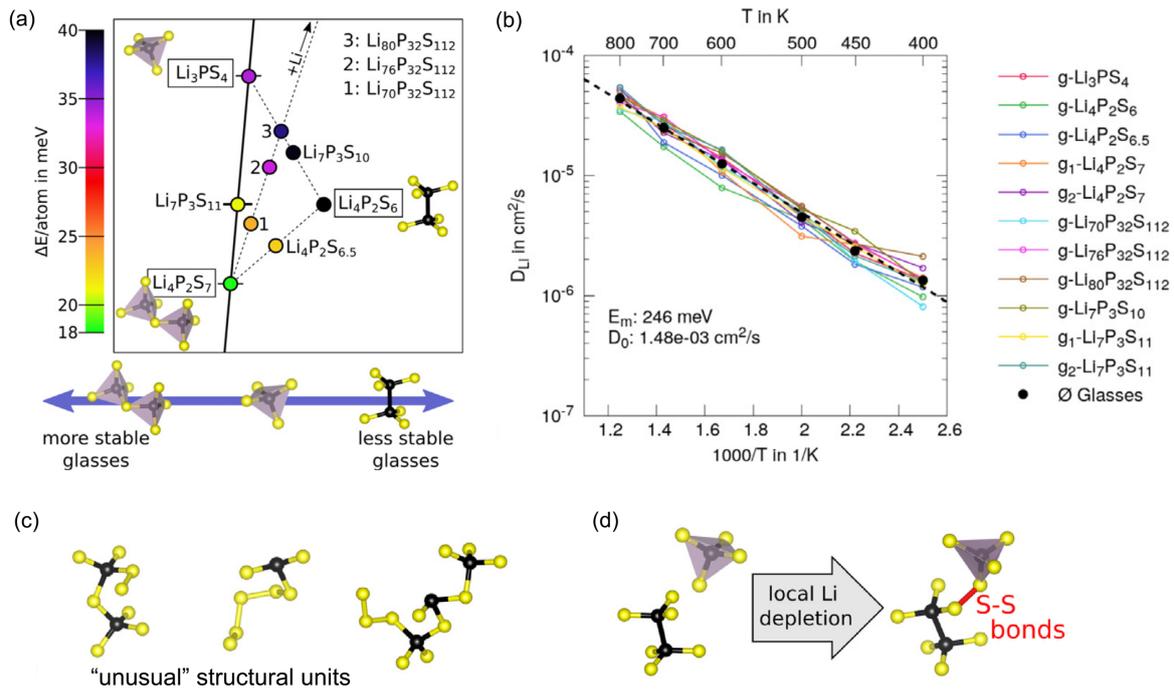

**Figure 4**. Thermodynamic stabilities, transport properties and local structures of $xLi_2S$-$(1-x)P_2S_5$ ($x$ = 0.67, 0.70 and 0.75) glasses. (a) Relative stabilities of glass phases against the three corner phases $Li_3PS_4$, $Li_4P_2S_7$ and $Li_4P_2S_6$; (b) Arrhenius plot of the studied glassy systems; (c) the ''unusual'' P-S structural units presented in the slowly quenched structure (at 10K/ps); and (d) the formation of S–S bonds at the lithium-deficient conditions in $Li_4P_2S_7$ glass[171]. (Copyright 2020, Elsevier)

Recently, J. Siegel et al. have performed AIMD simulations for $0.75Li_2S$-$0.25P_2S_5$ glass, and found the low-temperature paddle-wheel effect in this glassy solid electrolyte[95]. They performed AIMD simulation within the NPT ensemble for the $0.75Li_2S$-$0.25P_2S_5$ glass at T=300 K and P=1 bar, and eventually got a very low mass density of 1.56 g/cm$^3$. The most noteworthy aspect of their AIMD simulation is the observation of the paddle-wheel dynamics present in this glass at low temperatures. The coupled cation-anion dynamics enhance cation mobility. That is in $0.75Li_2S$-$0.25P_2S_5$ glass, lithium ion translational diffusion at RT via a correlated process (cation-cation interaction, multiple lithium ions simultaneous motion) dynamically coupling to the rotational motions of $PS_4^{3-}$ polyanions by the spatial, temporal, and energetic harmony. First, lithium ions diffusion remains coordinated to their neighboring $PS_4^{3-}$ tetrahedra by simultaneous reorientations of polyanions (see **Figure 3c**). Second, the power spectra obtained via the Fourier transform of the (angular) velocity autocorrelation function show a stronger overlap between the lithium vibrational and polyanion rotational modes than γ-$Li_3PS_4$ crystal (**Figure 3d**). Third, the activation energy barriers for lithium ion translation (0.22–0.25 eV) and anion rotation (0.27 eV) are close to each other. The low-density glasses containing simple polyanions and the polyanions exhibiting no long-range covalent network are expected to enhance the cation mobility even at low temperature by fostering paddle-wheel dynamics. Further discussions on the paddle-wheel effects in some superionic conductors are performed in the following Section 5.

Note that the calculated mass density at RT of 0.75$Li_2$S-0.25$P_2S_5$ glass is just 1.56 g/cm$^3$, much smaller than the experimental value of 1.93 g/cm$^3$ [35] and even the previous AIMD calculated density of 1.79 g/cm$^3$ [83]. Thereby, no wonder that the amorphous structure with such a small density would provide abundant free spaces for anion rotations even at RT. In addition, the calculated lithium ionic conductivities of this low-density 0.75$Li_2$S-0.25$P_2S_5$ glass are 7-19 mS/cm at RT, much higher than the previous AIMD calculated result of 0.088 mS/cm and even the experimental values of 0.28-0.32 mS/cm at RT[35, 172]. Therefore, the density issue of this AIMD simulation for 0.75$Li_2$S-0.25$P_2S_5$ glass should be careful and need further checking. The huge mass density difference between this AIMD simulation and the previous experiments may be because the pressure setting value of 1 bar for AIMD simulation is lower than those of the experiments during the mechanical milling process.

Very recently, Ohkubo et al. performed long AIMD runs (800 ps) at 300 K to study the $0.7Li_2S$-$0.3P_2S_5$ glass and its metastable crystal phase with the same composition, $Li_7P_3S_{11}$, promoting deeply understanding of the structural, dynamical, and electronic polarizability difference between the $0.7Li_2S$-$0.3P_2S_5$ glass and its metastable crystal[108]. Firstly, they analyzed the Li–Li correlation of $0.7Li_2S$-$0.3P_2S_5$ glass and $Li_7P_3S_{11}$ crystal by the Li–Li radial distribution functions, $g(r)$, and the running coordination number (see **Figure 5a**). No significant difference in the Li–Li $g(r)$ and its running coordination of the $0.7Li_2S$-$0.3P_2S_5$ glass vs. $Li_7P_3S_{11}$ crystal can be observed, and the extents of lithium ion collective motion in these two phases are almost equivalent. Therefore, the Li–Li arrangement and collective motion cannot satisfactorily explain the large difference in lithium ion conductivity between the $0.7Li_2S$-$0.3P_2S_5$ glass and $Li_7P_3S_{11}$ crystal. Although the excellent lithium ion conductivity of $Li_7P_3S_{11}$ crystal is usually considered to be associated with the lithium ion collective migration in the conduction paths. Additionally, the S–S $g(r)$ peak ranging from 3.5 to 4.0 Å indicates the intra-PS unit correlation directly shaping the lithium ion conduction path. The S–S $g(r)$ peak for the intra-PS units in $0.7Li_2S$-$0.3P_2S_5$ glass corresponds to a shorter distance than that of Li$_7$P$_3$S$_{11}$ crystal (**Figure 5b**), demonstrating the larger volumes for the lithium ion conduction paths in Li$_7$P$_3$S$_{11}$ crystal than 0.7Li$_2$S-0.3P$_2$S$_5$ glass.

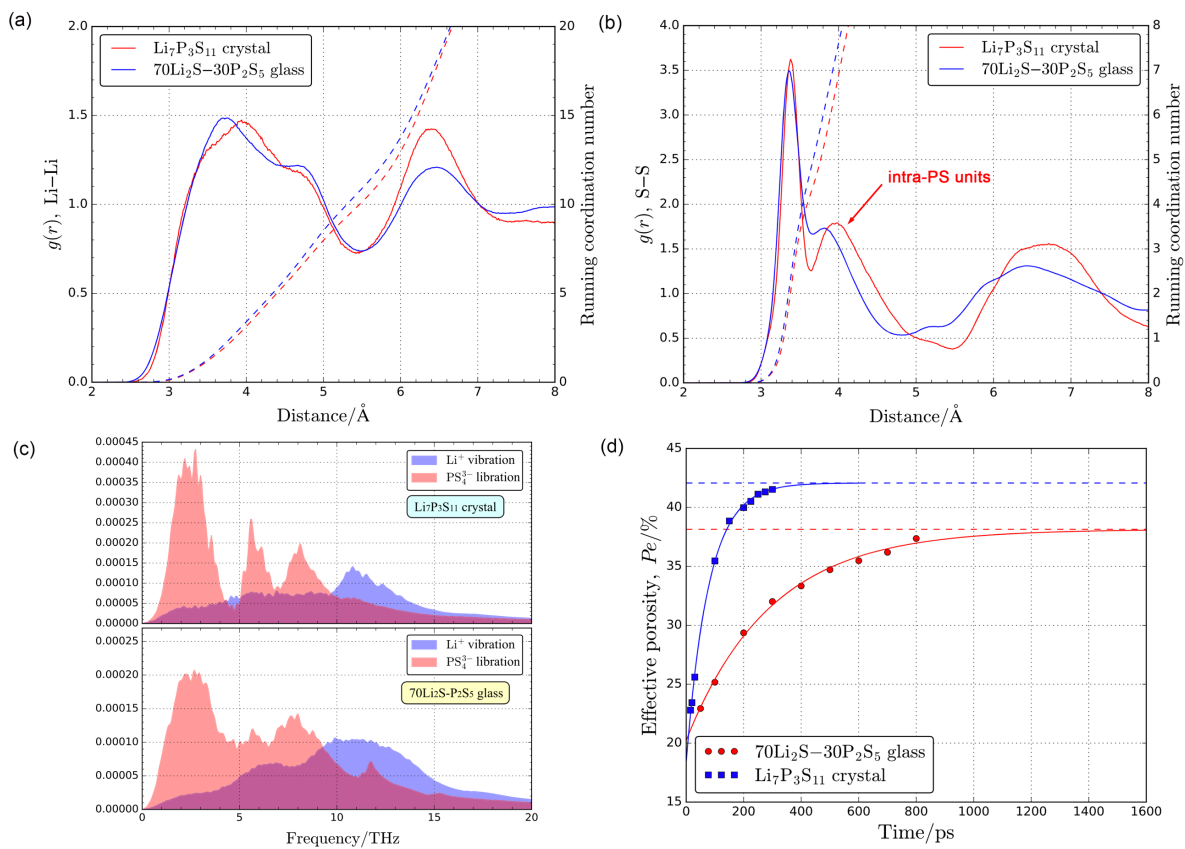

**Figure 5**. (a) Li-Li and (b) S–S radial distribution function (solid lines), g(r), and running coordination numbers (dashed lines) for 0.7Li$_2$S-0.3P$_2$S$_5$ glass and Li$_7$P$_3$S$_{11}$ crystal; (c) power spectra calculated from Fourier transform of the Li velocity autocorrelation and angular velocity autocorrelation functions of PS$_4^{3-}$ (d) time-dependent effective porosity for 0.7Li$_2$S-0.3P$_2$S$_5$ glass and Li$_7$P$_3$S$_{11}$ crystal. Parts (a-d) are reproduced with permission from reference[108], (Copyright 2020, American Chemical Society).

Previous researches have shown that the power spectrum obtained from the Fourier transform of (angular) velocity autocorrelation function can help us to assess the cooperative cation-anion motion qualitatively[95, 173]. However, the power spectra of Li$^+$ vibration and PS$_4^{3-}$ rotation in 0.7Li$_2$S-0.3P$_2$S$_5$ glass and Li$_7$P$_3$S$_{11}$ crystal show similar frequency modes and overlap (**Figure 5c**), suggesting that $0.7Li_2S-0.3P_2S_5$ glass and $Li_7P_3S_{11}$ crystal are characterized by similar coupling dynamics for $Li^+$ vibration and $PS_4^{3-}$ rotation. It is quite different from the recent study of comparing the $0.75Li_2S-0.25P_2S_5$ glass with a γ-$Li_3PS_4$ crystal, whose power spectra exhibit quite different overlaps between $Li^+$ vibration and the $PS_4^{3-}$ rotation modes[95]. This inconsistency is due to the very low $Li^+$ ionic conductivity of γ-$Li_3PS_4$, which is 3-4 orders of magnitude lower than that of $0.75Li_2S-0.25P_2S_5$ glass[14, 44, 172]. On the contrary, the ionic conductivity of $Li_7P_3S_{11}$ crystal is just 1-2 orders of magnitude higher than that of $0.7Li_2S-0.3P_2S_5$ glass[35, 172, 174]. Therefore, the power spectra of superionic conductors, e.g., $0.7Li_2S-0.3P_2S_5$ glass and $Li_7P_3S_{11}$ crystal, couldn't provide enough information for us to qualitatively evaluate the coupled motion. Even more exciting is that the simulation movies for AIMD production run of $0.7Li_2S-0.3P_2S_5$ glass provide direct evidences to assess the coupled cation-anion motion, which clearly show a frequently rotational motion of $PS_4^{3-}$ polyanion, while this rotational motion is not coupled with $Li^+$ translational motion. Moreover, $PS_4^{3-}$ rotational motion is not observed in $Li_7P_3S_{11}$ crystal. Therefore, both in $0.7Li_2S-0.3P_2S_5$ glass and $Li_7P_3S_{11}$ crystal, the occurrence probability of the coupled cation-anion dynamics is extremely low although they have vibration rotation mode overlaps. Further evidence should be explored to explain ionic conductivity difference between $0.7Li_2S-0.3P_2S_5$ glass and $Li_7P_3S_{11}$ crystal.

To characterize the total lithium ion diffusion space in $0.7Li_2S-0.3P_2S_5$ glass and $Li_7P_3S_{11}$ crystal, they analyzed the effective porosity for the AIMD simulation cells based on the lithium trajectories. The ion diffusion within the limited space in a cell is regarded as the effective porosity. The smaller effective porosity means more crowded ionic diffusion channels and inefficient ion diffusivity. **Figure 5d** shows that $Li_7P_3S_{11}$ crystal possesses a higher effective porosity than $0.7Li_2S$-$0.3P_2S_5$ glass. This reduction of the effective $Li^+$ diffusion space in $0.7Li_2S$-$0.3P_2S_5$ glass is attributed to the rotational motions of $PS_4^{3-}$ polyanions. Moreover, they also analyzed the sulfur polarizability by calculating the Born effective charge tensors, and found the uniformly anisotropic polarizability of sulfur ions in $0.7Li_2S$-$0.3P_2S_5$ glass is a characteristic property, which accounts for its lower lithiun ion diffusivity than $Li_7P_3S_{11}$ crystal[175]. Although the AIMD calculated ionic conductivity results is one orders of magnitude larger than that of the experimental data, this work of AIMD simulation aided comparing $0.7Li_2S$-$0.3P_2S_5$ glass with $Li_7P_3S_{11}$ crystal improves our understandings of the cation-anion coupled dynamics and the importance of anisotropic polarizability of anions for designing effective conduction paths in SSE materials.

For the classical MD simulations of $Li_2S$-$P_2S_5$ glass-ceramic, Kim et al. built a glass-interface-ceramic structural model (**Figure 6a**), and uncovered the differences of lithium-ion transport characteristics in $0.75Li_2S$-$0.25P_2S_5$ glass-ceramic (**Figure 6c**) [176]. In a MD model with the same composition ratio, lithium ions in glass region show highest ionic conductivities at RT, while crystalline γ-phase corresponds the lowest ionic conductivities (**Figure 6c**). lithium ion trajectory lines obtained from MD simulations (**Figure 6a**) clearly exhibit the differences of lithium ion diffusion in each region. To further analyze the region dependent lithium ion diffusion, they developed a new method to analyze the local structures of MD model, that is dividing the glass-interface-ceramic structure into three subsections and examines the S-sublattice in each local region. **Figure 6b** shows the local S-sublattice distributions in the region of glass-interface-ceramic structure. The S-sublattice of the crystalline γ-phase $Li_3PS_4$ has a hexagonally close-packed (HCP) pattern, while the S-sublattices of the glassy and interfacial structure show a mixed characteristic of HCP and cubic patterns. **Figure 6d** shows the variations of the S-sublattice type ratios along the z-axis direction from crystal to glass. The ratios of cubic S-sublattice grows rapidly in the interfacial region and approximately remain a constant in the glassy region. The approximately 40% ratio of cubic S-sublattice in the interfacial and glassy regions ensure relatively lower activation energy and higher ionic conductivities than that of crystalline γ-phase. As clarified, the glassy structures lacking long-range ordering have abundant atomic disorders and distortions, for example, lithium site disorders and $PS_4^{3-}$ polyanion reorientation in $0.75Li_2S-0.25P_2S_5$ glass. These disordered and distorted lithium ions, reorientated $PS_4^{3-}$ polyanions are dynamically coupled to lithium ion diffusion[95, 177], and would greatly reduce activation energy and improve lithium ionic conductivity. The glass-interface-ceramic MD results verify again that the arrangement and reorientation of polyanions significantly affects lithium ionic conductivity even for the same composition ratio.

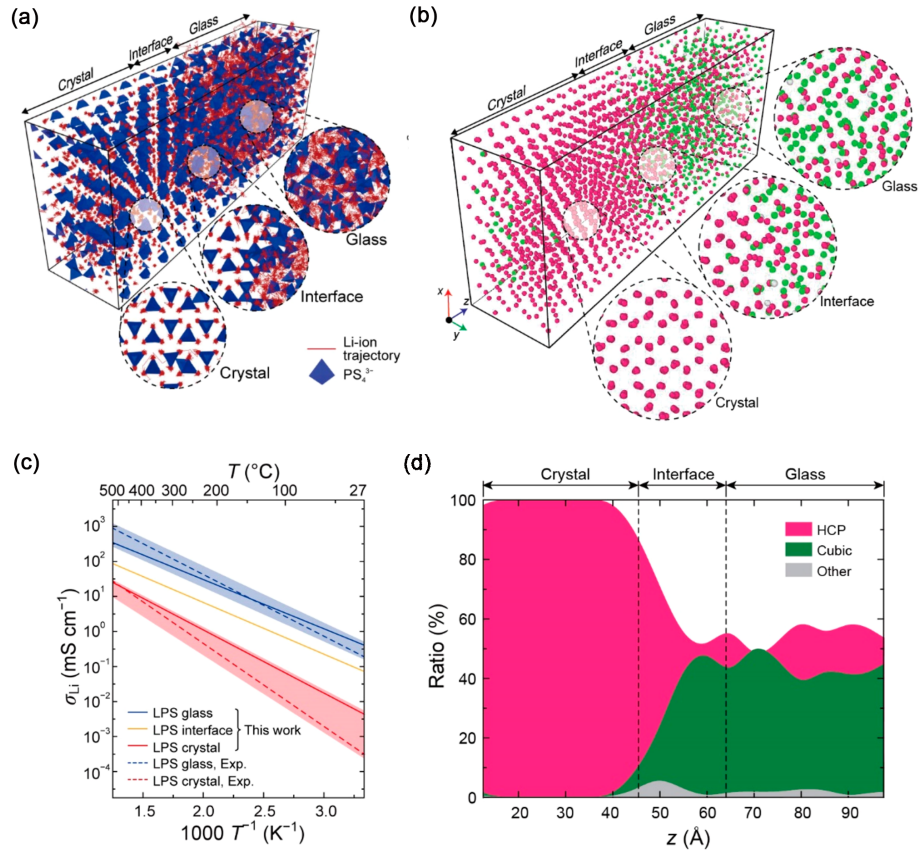

**Figure 6.** MD simulations of the glass-ceramic $0.75Li_2S-0.25P_2S_5$. (a) Trajectories of Li-ions in glass-ceramic structure at T=650 K for 500 ps, the red line indicates the Li-ion trajectory; (b) visual representation of S-sublattice distribution in the glass-ceramic LPS structure, the pink, green, and gray indicate the HCP, cubic, and other S-sublattices, respectively; (c) Arrhenius plots of the predicted lithium ionic conductivities of $0.75Li_2S-0.25P_2S_5$ glass, interface and crystal; and (d) variation of each S-sublattice ratio along the z-axis of structure model[176]. Parts (a-d) are reproduced with permission from reference, (Copyright 2019, American Chemical Society).

## 4. Sodium sulfide-type glasses

Since 2012, Hayashi et al. reported a $Na_3PS_4$ glass-ceramic with a good ionic conductivity of ~0.2 mS cm$^{-1}$ at 25 °C [178], many research efforts have attempted to further improve the ionic conductivity of $Na_3PS_4$ glass-ceramic through element doping[179-180] or forming composites[181-182]. Meanwhile, MD simulations were also performed to get a detailed understanding of the local atomic structures and their subsequent influences on the ionic conductivity of the experiment synthesized sodium-based glasses[183-184], and predict more new sodium-based glasses as the advanced electrolyte materials for solid-state sodium batteries[185].

Dive et al. utilized the combination of AIMD simulations with 144 atoms and the classical MD simulations with ~21000 atoms to study the effect of local structure on the Na$^+$ ion transport in the Na$_2$S-SiS$_2$ glass[183]. They found that the local short-range-order structures have great impacts on the distribution of stable Na$^+$ ion sites in the Na$_2$S-SiS$_2$ glass. Based on the local structure differences, three distinct hopping mechanisms for Na$^+$ ion migration in Na$_2$S-SiS$_2$ glass were identified (**Figure 7**), which significantly contributes to sodium ionic conductivity. That is the Na$^+$ local structures of initial state, transition state and final state have different coordination numbers of the neighboring S anions and SiS$_4$ tetrahedra. The type-I mechanism, the coordination of Na$^+$ ion with the neighboring S anions keeping constant throughout Na$^+$ ion hopping regardless of the coordination of SiS$_4$ tetrahedra (**Figure 7a**), corresponds to the smallest jump distance of 2.54 Å and the lowest activation energy barrier of 29.76 kJ/mol. In other words, the small change of Na$^+$ local structure during ion hopping would make low activation energy barrier for Na$^+$ ion migration, especially for the first coordination shell of neighboring S anions. Moreover, the classical MD simulations based on the empirical Buckingham potentials were performed to simulate much larger systems for accurately capturing the ionic conductivities of xNa$_2$S-(1-x)SiS$_2$ glasses and studying the effect of the composition change on ionic conductivity. An increase of Na$_2$S content (> 67%) significantly modifies the local structures of glasses, and creates more emergent secondary Na$^+$ ion sites, eventually leading to enhanced sodium ionic conductivity of glasses. This work widens our knowledge of the interactive relationship between Na$^+$ ion migration and the perturbation of Na$^+$ surrounding local structure in Na$_2$S-SiS$_2$ glasses.

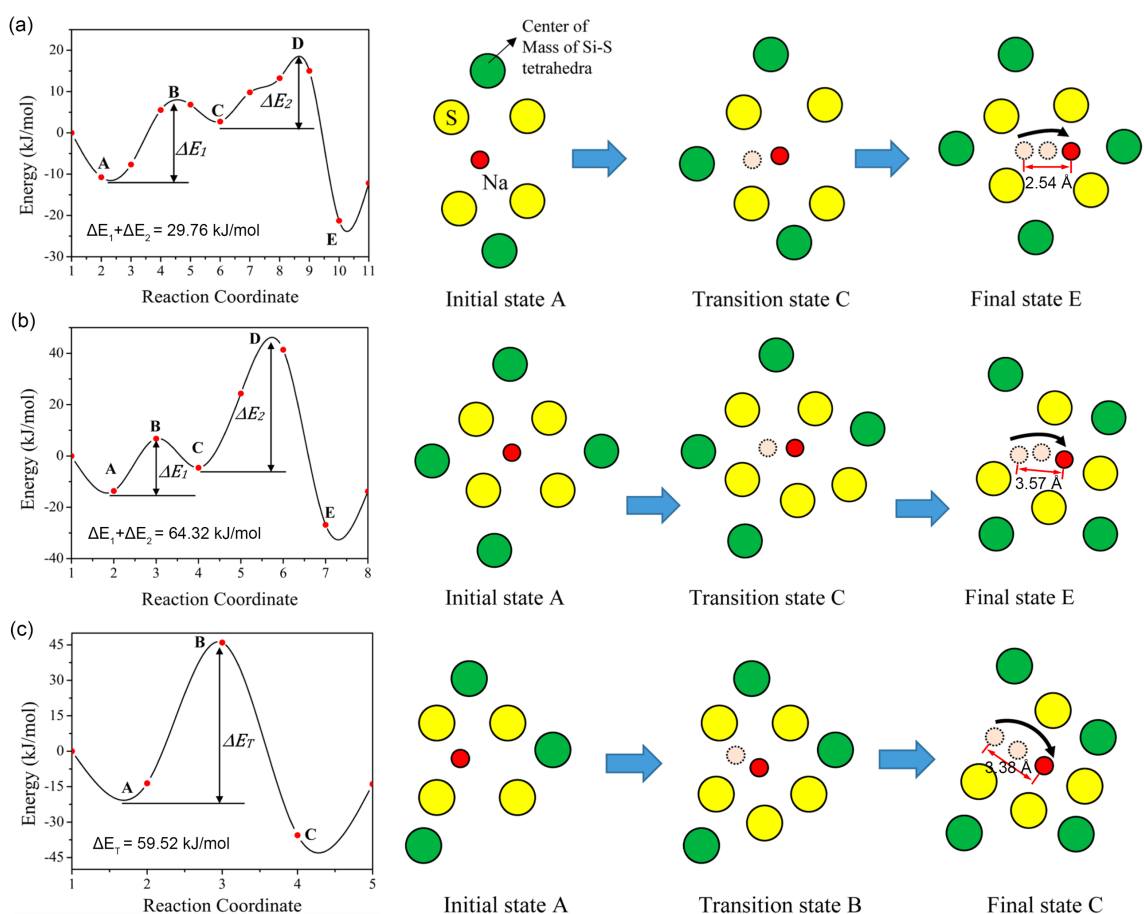

**Figure 7**. Three types hopping mechanism for Na$^+$ ion migration mechanism in Na$_2$S-SiS$_2$ glass, (a) type-I, (b) type-II, and (c) type-III. The green circles donate Si atoms, depicting the center of mass of SiS$_4$ tetrahedra, the yellow circles donate S atoms, the red bold circles donate the current site of Na ions, and the pink circles donate the prior occupied sites of Na ions. Parts (a-c) are reproduced with permission from reference[183], (Copyright 2018, American Chemical Society).

Another representative work of Dive et al. is AIMD simulation investigating the local structure changes, the formation of polysulfides, and their impact on Na$^+$ ionic conductivity of the Na$_2$S-P$_2$S$_5$ glasses with respect to different Na$_2$S contents[184]. In this work, the relative fractions of different polyanion groups (short-range-order structural units, **Figure 8a**) for a range of xNa$_2$S-(100-x)P$_2$S$_5$ glasses (33≤x≤75) were analyzed. Among these short-range-order polyanion groups, they found some polysulfide species acting as Na$^+$ ion trapping sites and thereby lowing sodium ionic conductivity. P$^{1P}$ and P$^{:3}$ polyanion groups are the major contributors to the formation of free sulfur and ultimately polysulfides in these glasses. With the increase of Na$_2$S content, the decreased relative fraction of P$^{1P}$ and P$^{:3}$ leads to low polysulfide concentration (**Figure 8b**) and low activation energy (**Figure 8c**). Hence, the decrease of polysulfide concentration together with decreased activation energy for Na$^+$ ion hopping would significantly improve sodium ionic conductivity of these glasses with higher Na$_2$S concentration (**Figure 8d**).

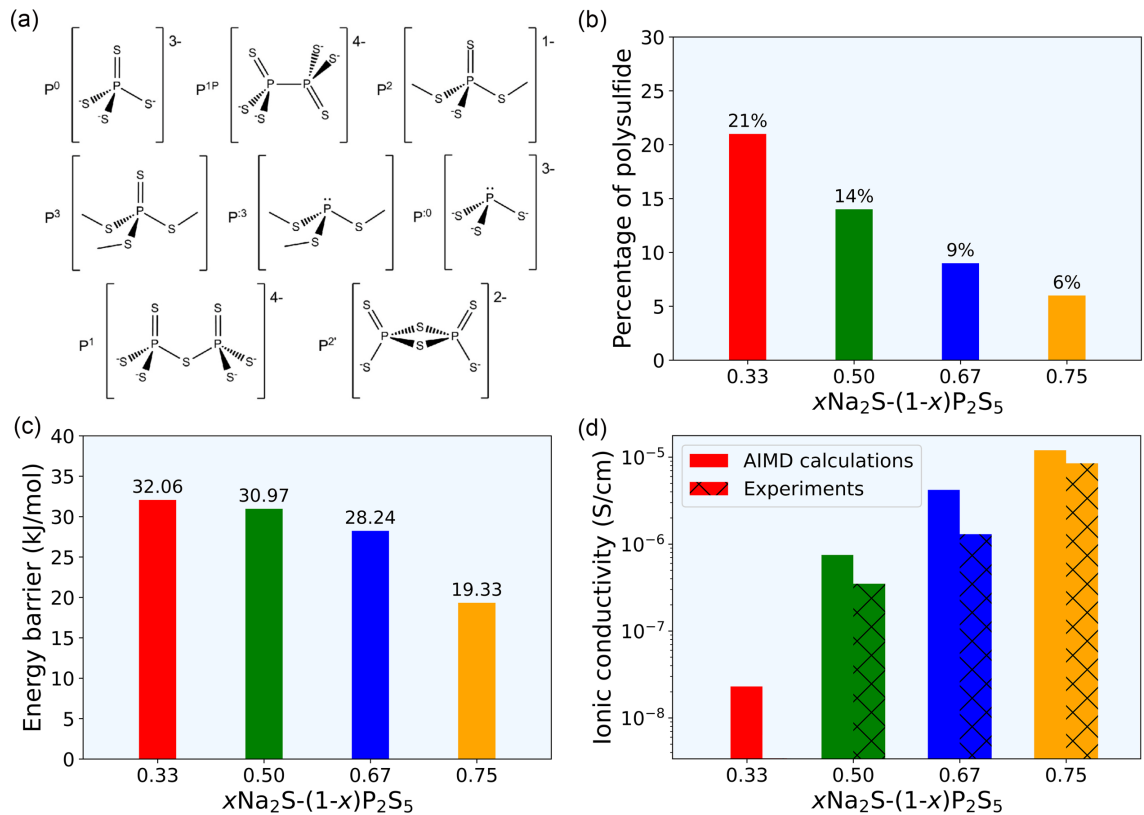

**Figure 8**. (a) Local short-range-order phosphorus environments present in xNa$_2$S-(100-x)P$_2$S$_5$ glasses; (b) Relative fractions of polysulfides; (c) calculated theoretical energy barriers for Na$^+$ ion hopping; (d) calculated ionic conductivity at 300 K compared with experimental measurements of the xNa$_2$S-(100-x)P$_2$S$_5$ glasses. Part (a) is reproduced with permission from references[184, 186], (Copyright 2019, Elsevier). Data of parts (b-d) are reproduced with permission from reference[184], (Copyright 2019, Elsevier).

## 5. Paddle-wheel effect

Early in 1972, a model of the connection between the high ionic conductivity and anion rotational disordering in α-Li$_2$SO$_4$ was proposed[187]. In this model, the tetrahedral SO$_4^{2-}$ sulfate groups would rotate, and widen low-energy migration passageways for lithium ion hopping, eventually enhancing lithium ionic conductivity[188]. These rotationally disordered behaviors of $SO_4^{2-}$ sulfate anions in α-$Li_2SO_4$ were later verified by the neutron diffraction experiments[189]. Similar cation–anion coupled interactions are also observed in other high-temperature sulfate and borohydride phases with fast ionic conductions, such as $LiNaSO_4$[190], $LiAgSO_4$[190], $LiBH_4$[191], $Li_2B_{12}H_{12}$[173] and $Na_2B_{12}H_{12}$[173, 192]. These pioneering studies show a strong dynamic coupling between the rotational motions of the sulfate and borohydride polyanions and the cation translational motions, called as the "paddle-wheel" mechanism[193].

As the paddle-wheel phenomenon is typically observed in high-temperature phases with large free volumes for polyanion rotation, exploiting the polyanion rotation in low-temperature materials present a great challenge. Fortunately, the polyanion rotation and even paddle-wheel effect at low-temperature in some glassy and crystalline superionic conductors have been recently observed by neutron diffraction experiments and AIMD simulations, such as $0.7Li_2S$-$0.3P_2S_5$[108], $0.75Li_2S$-$0.25P_2S_5$[95] glasses, $Li_{3.25}P_{0.75}Si_{0.25}S_4$[130], and $Na_{11}Sn_2PX_{12}$ (X = S and Se)[131] crystals, as summarized in Table 2. Enhancing anion rotational dynamics are coupled to and greatly constitute cation migration via the paddle-wheel mechanism (anion-cation interaction) by providing the driving force and widening the bottleneck size for cation diffusion along the lower energy barrier path [130-131]. The polyanion rotations in these two glasses of $0.7Li_2S$-$0.3P_2S_5$[108] and $0.75Li_2S$-$0.25P_2S_5$[95] make quite different contributions to total lithium ion diffusivity (Table 2). Therefore, the cooperative motions between cation vibration and polyanion rotation at the same time scale is very important for achieve excellent cation mobility, otherwise the polyanion rotation would reduce the effective space for cation migration and lower cation diffusivity. As show in **Figure 9,** these superionic conductors with low mass densities, simple polyanions and no long-range covalent network are expected to show the spatial, temporal, and energetic coupling anion-cation interplays at room temperature, eventually improving cation mobility. Enhancing anion-cation coupled dynamics serve as a complementary and general design principle for emerging fast ionic conductors with isolated polyanion groups, especially for the glassy systems. Note that sulfides are one of the most likely conductors, in which polyanion rotation can persist down to the room temperature, due to the larger polarizability of sulfides compared with oxides.

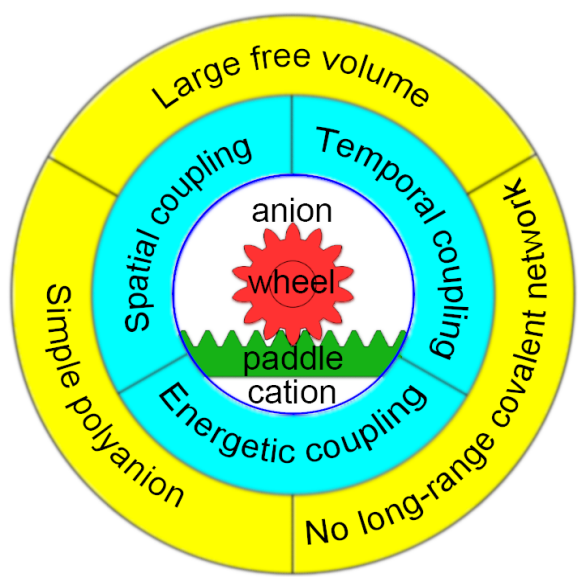

**Figure 9**. Schematic diagram of the requirements of local structure and coupling for the paddle-wheel effect presented in superionic conductors.

**Table 2**. Some reported superionic conductors with anion rotations at RT.

| System | Phase | Density determination method | AIMD temperature (K) | AIMD time (ps) | Anion rotation and cation transitional diffusion | Ref. |
|---|---|---|---|---|---|---|
| 0.75Li$_2$S-0.25P$_2$S$_5$ | glass | AIMD with NPT ensemble at 1 bar and 300 K | 300 | 300 | Coupled | 95 |
| 0.7Li$_2$S-0.3P$_2$S$_5$ | glass | Set to experimental value | 300 | 800 | Uncoupled | 108 |
| $\beta$-Li$_3$PS$_4$; Li$_{3.25}$Si$_{0.25}$P$_{0.75}$S$_4$ | crystal | DFT relaxation at 0 K | 1200 | 450 | Coupled | 130 |
| Na$_{11}$Sn$_2$PX$_{12}$ (X=S, Se) | crystal | DFT relaxation at 0 K | 1050 | 220 | Coupled | 131 |

## 6. Challenges in glass MD simulation

In spite of rapid progresses and increasing applications of MD simulation on studying the structures, properties, and structure-property correlations of glassy SSE materials, MD simulations still encounter many challenges in studying the multicomponent glass and amorphous materials[77]. For example, for the AIMD calculations of $Li_2S$–$P_2S_5$ binary glass, and the three calculated results of lithium ionic conductivity at RT are quite different from each other (Table 3), and are also different from the experimental values of 0.75-1.10 mS/cm[36, 194]. At present, modeling glass structure mainly adopts the melt-quench method based on MD simulation. Since the glass structures finally produced by the melt-quench path are not unique and have a certain randomness, the calculated ionic conductivity values from MD simulations are also unrepeatable to a certain extent.

Our previous research found that for $0.75Li_2S$–$0.25P_2S_5$ ($Li_3PS_4$) glass, even using the same initial $0.75Li_2S$–$0.25P_2S_5$ glass structure for AIMD simulations, the calculated results of activation energy barrier of different repetitions are also quite different from each other (**Figure 10**). The reproducibility of calculated activation energy barrier is very poor, and the calculation deviation of the activation energy barrier is as high as 60 meV, so the corresponding lithium ionic conductivities at RT would differ by 1-2 orders of magnitude. However, the calculated activation energy barriers from AIMD simulations of the crystalline $Li_3PS_4$ (β and γ phases) have relatively good repeatability and are basically consistent with the previously reported calculation results of $Li_3PS_4$[195].

Therefore, MD calculated transport properties such as ionic conductivity of glassy solid electrolytes are much sensitive and demanding to the initial glass structure and its modeling process. Compared to the crystalline materials, MD simulation processes of glassy materials are more complicated, and the simulation time and cost are also higher. So far, the initial modeling structures of multi-glassy solid electrolytes and their MD simulations are still huge challenges. These challenges include but are not limited to the three main parts: the reliable and transferable potentials models and parameters for multicomponent systems, the cooling rate effect during quench process for glass formation, and the simulation box size effect on certain properties.

**Table 3**. Important parameters, calculated activation energy barrier ($E_a$) and ionic conductivity at RT of 75%$Li_2S$–25%$P_2S_5$ glass from AIMD simulations

| Ensemble | Heat bath method | Simulation time (ps) | density (g/cm$^3$) | Number of atoms | $E_a$ (eV) | $\sigma_{300K}$ (mS·cm$^{-1}$) | Ref |
|---|---|---|---|---|---|---|---|
| NVT | Nose | - | 1.73 | 256 | 0.26 | 10.4 | 171 |
| NPT | Langevin | 80 | 1.56 | 160 | 0.22 | 19 | 95 |
| NPT | Nose | 30-40 | 1.80 | 130 | - | 0.088 | 83 |

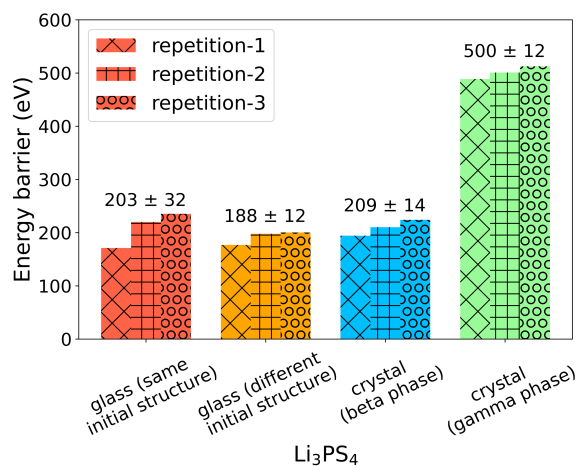

**Figure 10.** Comparisons of the repetitive calculated results of activation energy barriers for lithium ion diffusion in crystalline $Li_3PS_4$ and glassy 75%$Li_2S$–25%$P_2S_5$

## 6.1 Empirical potential

The first challenge is the issue of empirical potential for MD simulations. Reliable, transferable, and efficient potentials determine the accuracy and validity of the final simulated structures and properties, especially for the dynamic properties, which are more sensitive to the potentials. In practice, the availability of potential models is usually the limiting factor whether a system can be simulated. As most of the early MD simulations of glassy SSE materials focused on simpler binary glass systems, e.g., $Li_2S$-$P_2S_5$, the available potentials in the literature are also limited to these simple components and very few potential sets are applicable to multicomponent systems[196]. In addition, most potential parameters were fitted in the crystalline systems based on the structure and mechanical properties under ambient temperature and even DFT ground state (0 K). Therefore, temperature dependent properties, such as melting temperature, thermal expansion coefficient and heat capacity, are not commonly used in the fitting potential parameters. Conversely, AIMD simulations obtaining the interatomic forces from first principles calculations and being free from any parametric force field are the ultimate solutions to the potential issue. In addition, AIMD simulations can effectively capture the polarizable nature interatomic forces important for ionic diffusion[185]. However, at the moment, AIMD is still limited to the small simulation system size and relatively short times (up to a few hundred atoms and tens of picoseconds) due to high computational cost of first principles calculations. It is expected that AIMD will be applied to more and more multicomponent glassy SSE materials due to availability of high-performance computing facilities but classical MD will not likely to be fully replaced by it in the near foreseeable future.

## 6.2  Cooling rate

Another big challenge is the cooling rate of the quench process for glass formation in MD simulations. To guarantee the integration accuracy of motion equation, MD simulations use the time step in the order of femtosecond (fs) as same as particle motion, which limits the total accessible simulation time ranging from hundreds pico-seconds (ps) to a few nano-seconds (ns), and corresponds to the cooling rate at K/ps level during the simulated melt-and-quench process[77, 104]. Thus, the used cooling rate in MD simulations are many orders of magnitude larger than the typically achieved in experiments at 1–100 K/s level, limited by the accessible MD simulation time. Now, direct comparison of cooling rate values from experiment and MD simulations is meaningless, and there is a large gap of cooling rate between the MD simulations and experiments. It's still one of the common criticisms of MD simulations of glass from the viewpoints of our experimentalists.

Sometimes, the cooling rate has great impacts on the MD simulation quenched amorphous structures, especially for the medium-range order structures, and their relevant property features[197-198]. Too larger cooling rate would cause the insufficient relaxation for local structures with residual stress and the formation of unrealistic structural defects[199], while too smaller cooling rate would lead to longer quench times, consuming large amounts of computing resources. The most suitable cooling rate for your amorphous structure should be specifically and carefully investigated with keeping the balance between the accuracy of your concerned properties and the affordable computing resource.

## 6.3 Simulation cell size

Although applying the periodic boundary condition in MD simulations, the number of atoms in glass classical MD and AIMD simulations tens of thousands and several hundreds, respectively, due to the limited computational costs, which are many orders of magnitude smaller than the real glass samples. For the $Li_7La_3Zr_2O_{12}$ crystalline superionic conductor, it has been reported that its dynamical properties of self-diffusivity, ionic conductivity, and Haven ratio have a relatively weak dependence on the simulation cell size[200]. While for many MD simulations of organic liquids and glassy materials[201-202], it is observed that the structural and dynamical properties obtained from MD results are much sensitive to the simulation cell size. For example, the atomic self-diffusivity is often found to increase with increasing cell size[203], and the clustering behavior of low concentration components of rare earth oxides in silica glasses should use sufficiently large cell size in order to get statistically meaningful results[204]. The small cell size also limits the accuracy of describing the variations of structure and property with respect to different glass compositions[183]. In addition, according to our recent work, the calculated densities of the melt-quench MD simulation determined amorphous structures show great dependence on the cell size, which would eventually affect the calculated dynamical properties significantly. Therefore, the cell size effect may be an issue in our MD simulations of glassy SSE materials, and it need to be carefully checked.

## 7. Conclusion and Perspectives

In this review, we present a summary of the common computational methods utilized for studying the amorphous inorganic materials, review the applications of these computational methods for the study of lithium and sodium sulfide-type glasses for solid-state batteries, and provided perspectives on the challenges and potential future developments in the computational studying on the new glassy-state SSE materials. Although there are some reported glassy SSE materials, it is conceivable that there are some chemical spaces for exploring the new glasses with the elementary composition of A-B-C-D (A = Li, Na and K, B = S, Se and Te, C is P, and D = Cl, Br and I), including new $Li_2S$-based glasses, $Na_2S$-based glasses, and even the dual-ion $Li_2S$-$Na_2S$ based glasses. The material computational researchers are amenable to screen and discover new glassy SSE materials in the whole chemical space, and fully figure out the relationships among the structure, composition and ion diffusion property of glassy SSE materials, especially for paying attention to "paddle-wheel effect".

On one hand, the data-driven machine learning should be actively employed for analyzing the structural and dynamical properties of new glassy SSE materials from the high-throughput calculations by using the Mpmorph workflow. On the other hand, developing the machine learning interatomic potential for classical MD simulations and utilize the machine learning-driven AIMD simulation to speed up the computational research of new glassy SSE materials, for example, we can utilize the combination of a genetic algorithm and a specialized machine learning potential to speed up the sampling of the low-energy atomic configurations in the entire amorphous $x$$Li_2S$-$y$$P_2S_5$-$(1$-$x$-$y)$ LiI phase space, and build a universal machine learning interatomic potential for the long-time and large-scale MD simulations of the $x$$Li_2S$-$y$$P_2S_5$-$(1$-$x$-$y)$ LiI glass systems with various molar ratios. Last, the increasingly normal computing power and faster algorithms would help to bring simulation time up from picoseconds to microseconds for the glassy SSE materials with thousands of atoms in cell. It is conceivable that, with ever-increasing high-performance computational facilities and development of methodologies and new machine learning interatomic potentials, MD simulation of glassy SSE materials is no longer a challenge and will play more and more important roles in the future researches of glassy SSEs. We hope this review could facilitate and accelerate the future computational screening and discovering more glassy-state SSE materials for the solid-state batteries.

## Acknowledgements

This work is supported by the Talent Research Startup Funds of Nanjing University of Aeronautics and Astronautics (1006-YAH21005).